\documentclass[tighten,iop,floatfix]{emulateapj}
\pdfoutput=1

\usepackage{epsfig}
\usepackage{amsmath}
\usepackage{natbib}
\usepackage[OT4]{fontenc}
\usepackage{graphicx}
\usepackage{subfigure}

\usepackage{hyperref}
\hypersetup{colorlinks=true,citecolor=blue}
\usepackage{breakurl}

\newcommand{\ANNz}{ANN$z$}
\newcommand{\ttt}{\texttt}
\newcommand{\ti}{$\times$}

\shorttitle{2MPZ: comprehensive 3D census of the whole sky}
\shortauthors{Bilicki, Jarrett, Peacock, Cluver \& Steward}

\submitted{Accepted to the Astrophysical Journal Supplement}

\begin{document}

\title{2MASS Photometric Redshift catalog:\\a comprehensive three-dimensional census of the whole sky}

\author{Maciej Bilicki$^{1,2,\ast}$, Thomas H.\ Jarrett$^1$, John A.\ Peacock$^3$, Michelle E.\ Cluver$^{1,4}$ and Louise Steward$^1$}
\affil{$^1$Astrophysics, Cosmology and Gravity Centre, Department of Astronomy, University of Cape Town, Rondebosch, South Africa\\
$^2$Kepler Institute of Astronomy, University of Zielona G\'{o}ra, ul.\ Lubuska 2, 65-265 Zielona G\'{o}ra, Poland\\
$^3$Institute for Astronomy, University of Edinburgh, Royal Observatory, Edinburgh EH9 3HJ, UK\\
$^4$Australian Astronomical Observatory, P.O. Box 915, North Ryde, NSW 1670, Australia\\
$^\ast$Email: maciek(at)ast.uct.ac.za}

\begin{abstract}
Key cosmological applications require the three-dimensional galaxy distribution on the entire celestial sphere. These include measuring the gravitational pull on the Local Group, estimating the large-scale bulk flow and testing the Copernican principle. However, the largest all-sky redshift surveys -- the 2MRS and IRAS PSC$z$ -- have median redshifts of  only $z=0.03$ and sample the very local Universe. There exist all-sky galaxy catalogs reaching much deeper -- SuperCOSMOS in the optical, 2MASS in the near-IR and WISE in the mid-IR -- but these lack complete redshift information. At present, the only rapid way towards larger 3D catalogs covering the whole sky is through photometric redshift techniques. In this paper we present the 2MASS Photometric Redshift catalog (2MPZ) containing 1 million galaxies, constructed by cross-matching 2MASS XSC, WISE and SuperCOSMOS all-sky samples and employing the artificial neural network approach (the \ANNz\ algorithm), trained on such redshift surveys as SDSS, 6dFGS and 2dFGRS.  The derived photometric redshifts have  errors nearly independent of distance, with an all-sky   accuracy of $\sigma_z = 0.015$ and a very small percentage of outliers.  In this way, we obtain redshift estimates with a   typical precision of 12\% for all the 2MASS XSC galaxies that lack   spectroscopy. In addition, we have made an early effort towards probing the entire 3D sky beyond 2MASS, by pairing up WISE with SuperCOSMOS and training the \ANNz\ on GAMA redshift data reaching currently to $z_\mathrm{med}\sim 0.2$. This has yielded photo-$z$ accuracies comparable to those in the 2MPZ. These all-sky photo-$z$ catalogs, with a median $z\sim0.1$ for the 2MPZ, and significantly deeper for future WISE-based samples, will be the largest and most complete of their kind for the foreseeable future.
\end{abstract}

%
%
%
\section{Introduction}
\label{Sec: Introduction} 

Observational studies of the large-scale structure in the Universe have made great progress in recent years through such projects as the 2dF Galaxy Redshift Survey (\mbox{2dFGRS}, \citealt{2dF}), the Sloan Digital Sky Survey (SDSS, \citealt{SDSS.III}) and the 6dF Galaxy Survey (6dFGS, \citealt{6dF}). These and other current and near-future redshift surveys have been highly successful in extracting cosmological information from the properties of the cosmic web.

\begin{figure}[!t]
\centering
\includegraphics[width=0.48\textwidth]{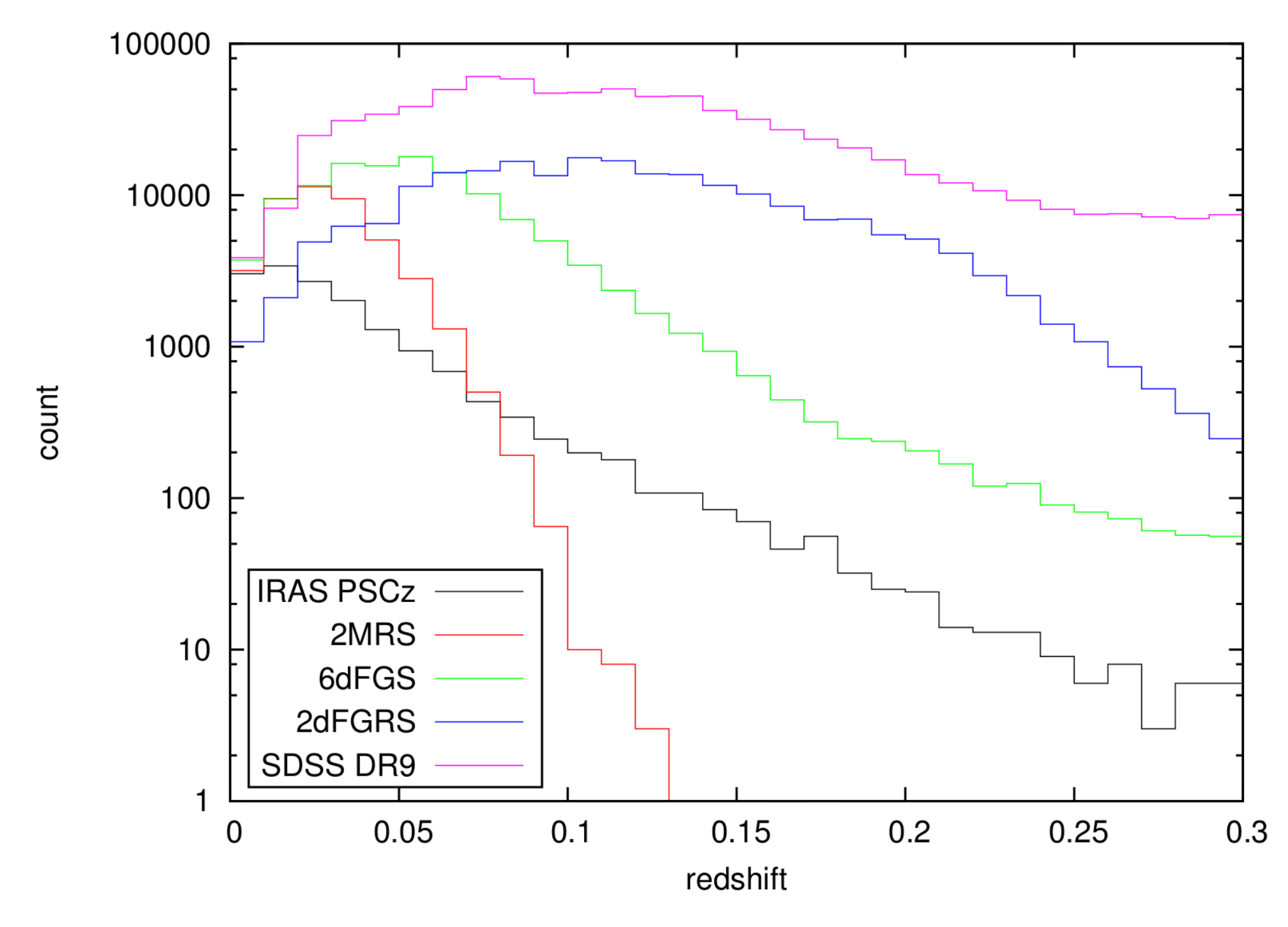}
\caption{Redshift distribution of the most important galaxy surveys of   the low-redshift Universe. Only the two shallowest of these, namely   2MRS and PSC$z$, cover (almost) the entire sky. Note the     logarithmic scaling of the $y$-axis.}
\label{Fig: Survey Histograms}
\end{figure}

Despite these achievements, some aspects of large-scale structure studies still require major improvements on the observational side. In particular, we  lack a dense, deep and uniform three-dimensional catalog of galaxies that covers the whole sky. Currently the two largest all-sky catalogs of this type are the IRAS Point Source Catalog Redshift Survey \citep[PSCz,][]{PSCz} and the 2MASS Redshift Survey \citep[2MRS,][]{2MRS}, based respectively on the observations made with the Infrared Astronomical Satellite \citep{IRAS} and the ground-based Two Micron All Sky Survey \citep{2MASS}.  The PSC$z$ contains more than 15,000 IRAS-selected galaxies (with fluxes larger than 0.6~Jy in the 60~$\mu$m band), covers 84\% of the sky and has a median redshift of $z=0.026$. The 2MRS is larger and denser, but slightly shallower (in terms of an effective depth), containing over 43,000 galaxies with $K_s$-band magnitudes below $11.75$, distributed over 91\% of the sky and with a median redshift of $z=0.028$. The number of galaxies and the depth of these surveys are an order of magnitude smaller than in the currently largest three-dimensional low-redshift surveys, such as the 2dFGRS or SDSS; see table 1 in \cite{2MRS} for a comparison of recent and historic catalogs, and Figure~\ref{Fig: Survey Histograms} herein for redshift distributions of several surveys of the nearby Universe.  An important additional remark is that even these `all-sky' surveys are highly incomplete for the swath of the sky that is obscured by the foreground Milky Way, the so-called Zone of Avoidance (ZoA). In the case of the near-infrared  2MRS,  for which the Galactic extinction is not as significant as in the optical, it is source confusion and stellar crowding that limits the detection of galaxies behind the Milky Way \citep{Jarr00ZoA,KKJ05}.

Comprehensive all-sky redshift surveys are challenging: a sample that covers the \textit{entire} celestial sphere and remain uniform in terms of photometry and other observable properties requires either a single space-borne instrument or twin ground-based telescopes. Converting this into a \textit{three-dimensional} all-sky sample of galaxies requires subsequent cross-matching with existing redshift surveys and spectroscopic follow-up for those sources that have no redshift information (see e.g.\ \citealt{2MRS}). Currently no such all-sky redshift surveys in the optical or infrared bands are being performed or even envisaged: for instance, the forthcoming major cosmological space mission of the European Space Agency, Euclid \citep{Euclid}, which aims at imaging a billion galaxies and measuring nearly 50 million galaxy redshifts, will cover $\sim$15,000 deg$^2$. Another promising concept, known as TAIPAN \citep{CBB13}, will follow up the \mbox{6dFGS} using the UK Schmidt Telescope, with a 2-magnitude deeper redshift survey matched to the catalogs of WISE and VISTA-VHS; but again, would not cover the entire sky, unless it is extended to the North, with the use of an additional instrument. Thus sadly the PSC$z$ and 2MRS are likely to remain the deepest complete catalogs of their kind for many years. Beyond the optical and infrared domains, the future radio surveys show promise to achieve all-sky redshift coverage: the WALLABY $HI$ survey, planned for the Australian SKA Pathfinder (ASKAP, \citealt{ASKAP}), and its Northern counterpart, APERTIF \citep{APERTIF} using the Westerbork Radio Telescope in the Netherlands.

The principal aim of the present paper is to show that by combining the largest \textit{photometric} all-sky galaxy catalogs we obtain reliable and unbiased photometric redshifts over the entire celestial sphere, reaching currently a median $z\sim0.1$ and at least double this in the near future, with reasonably small and well-behaved errors. The key is to use uniform photometric measurements spanning from the optical to the mid-infrared. Such data are available through 2MASS, WISE and SuperCOSMOS all-sky samples. Of these three surveys, 2MASS is the shallowest, so accordingly we based our analysis on the 2MASS Galaxy Catalog \citep{2MASSXSC}. Thanks to the multi-wavelength coverage, we were able to efficiently apply the artificial neural network approach (the \ANNz\ algorithm), trained on such redshift surveys as discussed above and thoroughly tested in selected fields, finally constructing the 2MASS Photometric Redshift catalog (2MPZ). In addition, we briefly consider the prospects of probing deeper than allowed by the 2MASS sample, by using a combination of WISE and optical imaging.

The paper is organized as follows. In Section \ref{Sec: Need for all-sky}  we present the motivations for obtaining deeper all-sky redshift surveys. Section \ref{Sec: All-sky photo-z} summarizes efforts to date regarding all-sky photometric redshift catalogs. In Section \ref{Sec: Photometry and redshifts} we describe the photometric and redshift surveys used within our project. A discussion on the photometric redshift methods, including the one we adopted here, is given in Section \ref{Sec: Photo-z framework}. Section \ref{Sec: Photo-z tests} presents tests of photometric redshift estimation in selected fields. A description of the all-sky 2MASS Photometric Redshift catalog follows in Section \ref{Sec: 2MPZ}.  We conclude and present future prospects in Section \ref{Sec: Summary}.

\section{The need for all-sky galaxy surveys}
\label{Sec: Need for all-sky}

Several cosmological problems call for all-sky three-dimensional surveys of at least a comparable depth to that of the SDSS main galaxy sample or 2dFGRS. Firstly, with the advent of precision cosmology it has become important to probe \textit{observationally} the degree of the homogeneity and isotropy of the late Universe \citep{Clarkson12}, as these two properties, the Copernican Principle (CP), underlie the standard cosmological model. Although tests applicable to surveys of even limited sky coverage exist \citep[e.g.][]{Hogg05,Sarkar09,GH12,Marinoni12,Scrimgeour12}, their usefulness for all-sky redshift surveys is currently limited by the shallowness of the latter.

Secondly, some important cosmological studies -- many of which also probe the CP in the low-redshift Universe, even if indirectly -- can be performed only with the use of an all-sky redshift survey. For instance, such a catalog is essential to examine the amplitude and the scale of convergence of the gravitational acceleration of the Local Group, via the so-called clustering dipole in galaxy distribution \citep[e.g.][and references therein]{BCJM11}. Although this measurement is possible, in principle, with the use of a `2.5-dimensional' catalog (i.e.\ with galaxy positions and fluxes only), redshift information is then essential for determination and characterization of the attractors and vital when comparing with theoretical expectations of particular cosmological models \citep{JVW90, LKH90}. The issue of the dipole convergence or lack thereof, as well as of the sources of the pull on the Local Group, still remains contentious, due to the lack of sufficiently deep and dense all-sky redshift surveys (see e.g.\ \citealt{BP06,Erdogdu06,KE06,Lavaux10}), as well as paucity of information in the ZoA, where many attractors are known or suspected to lie \citep[e.g.][]{Norma}.

Next, as was shown by \cite{NBD11} and more recently applied by \cite{BDN12}, an all-sky redshift catalog with uniform photometry can be used to place constraints on the amplitude and direction of the large-scale bulk flow, i.e.\ net peculiar motions of galaxies in a sphere centered on the Local Group, on scales larger than currently available. Such a measurement traditionally requires catalogs of peculiar velocities \citep[see e.g.][for recent   contributions]{ND11,Turnbull12}, which are both limited to the very local Universe ($z < 0.02$) and compromised by large uncertainties in distance indicators used. In order to probe beyond the reach of peculiar velocity catalogs, i.e.\ $\gtrsim100$~Mpc, galaxy clusters are being used, notably through the kinematic Sunyaev-Zel'dovich effect \citep[kSZ,][]{kSZ}. Nevertheless, the attempts to measure this effect still yield relatively low signal-to-noise, resulting in conflicting estimates of the bulk flows from this method, including claims of excessively high amplitudes on very large scales \citep[e.g.][]{KABKE,PlanckBulkFlow}.

A similar idea to the one discussed above lies behind the method of \cite{NBD12} to probe the growth rate parameter $f$, combined with the linear bias $b$ into $\beta \equiv f(\Omega) / b$, at the effective redshift of a survey, independently of redshift-space distortions \citep{Beutler12} or velocity-velocity comparisons \citep{Davis11}.  Using $\sim$45,000 2MRS galaxies, \cite{BDN12} were able to obtain $\sim25$\% accuracy on $\beta$; this error should decline as $N^{-1/2}$ for larger samples. Again, uniform and accurate photometry, large number of galaxies and wide-angle sky coverage are crucial for this approach.

Finally, the widest possible angular coverage of the redshift survey is desirable in those analyses where cross-correlation between the low-redshift large-scale structure (LSS) and cosmic microwave background (CMB) temperature distribution is explored. The most widely studied in this respect is the integrated Sachs-Wolfe effect \citep[ISW,][]{ISW} -- an important cosmological probe, as its detection gives confirmation of dark energy in the late Universe. The all-sky surveys used for the ISW measurements include 2MASS \citep{ALS04,Rassat07,FP10} and the Wide-field Infrared Survey Explorer \citep[WISE,][]{Kovacs13}; but the significance of ISW detection from these and other analyses is still weak, and deeper wide-angle redshift surveys would be valuable. Another possible application of an all-sky redshift survey is the gravitational lensing of the CMB by the LSS; see for instance the recent analysis by the Planck team \citep{PlanckGravLens} and the strong cross-correlation with WISE quasars measured by \cite{Geach13}.

\section{All-sky photometric redshifts}
\label{Sec: All-sky photo-z}

Currently the only rapid route to an all-sky three-dimensional catalog that would be significantly deeper than 2MRS or PSC$z$ is to use a broad-band photometric survey, such as 2MASS, gather all available spectroscopic redshifts for its galaxies (from other surveys) and estimate  \textit{photometric redshifts} for the remaining objects.  The technique of photometric redshifts \citep[e.g.][and references therein]{Benitez00,BMP00} is now widely used: e.g., the MegaZ-LRG sample of $\sim$1.5 million luminous red galaxies from the SDSS DR6 \citep{ABLR11} is one of many prominent examples of a successful application.

In the present paper we describe our successful efforts towards obtaining the deepest three-dimensional catalog of galaxies covering the entire sky, based mainly on 2MASS and its near-infrared measurements of resolved sources. Using this sample for such a purpose is not a new idea: almost a decade ago \cite{XSCz} used the three-band near-infrared photometry of 2MASS to estimate in a simple way luminosity distances to galaxies (the 2MASS XSC$z$). This method assumes that all galaxies have similar luminosities ($L^*$) and what modifies their near-infrared colors is the cosmic reddening. The measured integrated flux was the primary component, while the near-infrared colors added secondary information. This technique, being very crude in terms of accuracy (20 to 30\%), still provided a means to generate qualitative maps of the spatial distribution of galaxies and thereby construct an all-sky `big picture' view of the local Universe\footnote{See   e.g.\ \url{http://www.ast.uct.ac.za/\~jarrett/lss/} for some   whole-sky visualizations.}.

The XSCz method was particularly suitable to reveal clusters of galaxies as the redshift uncertainty of this approach declines with the square root of the number of cluster members seen by 2MASS. The photometric redshifts of galaxy clusters as estimated by \cite{XSCz} using only 2MASS photometry were typically accurate to 20\%, or worse for individual sources, and it was concluded that near-infrared colors alone are not sufficient as a discriminant of galaxy redshifts.  Such photo-$z$'z are ultimately constrained by the fact that the 2MASS $J$, $H$ and $K_s$ bands are, on the one hand, notably sensitive to the photospherically emitted light from evolved stars (e.g., K giants), and hence able to track the cosmological $k$-correction of this light out to low ($z < 0.2$) redshifts, but on the other hand insensitive to star forming, low-surface brightness and dwarf galaxies, resulting in color redundancies that are difficult to track with redshift. As we emphasize in the present paper, the solution is to add more photometric bands, at least one in the optical and one in the mid-infrared.

Another approach to obtain an all-sky 3D catalog based on 2MASS was taken by \cite{FP10}. Working in the context of analyzing the ISW signal, they derived photometric redshift estimates by matching the 2MASS data with optical catalogs generated from SuperCOSMOS scans of major photographic sky surveys \citep{SCOS1}.  A five-band $BRJHK$ photometric data set was used, calibrated via spectroscopic redshifts for about 30\% of the galaxies, thanks to SDSS, 2dFGRS and 6dFGS. This yielded photometric redshifts with an rms dispersion of about 0.033. Even better results can be achieved by adding further optical/UV photometry such as the SDSS \citep{Wang08,WHG09} or additionally the Galaxy Evolution Explorer \citep[GALEX,][]{WS06,Way09}, although this is only possible over a fraction of the sky.

In this present work, we show that by combining the 2MASS XSC with optical data from SuperCOSMOS and mid-infrared from WISE, and using the powerful machine-learning neural network method, photometric redshifts are very significantly ameliorated, to an rms error of around 0.015, and any overall biases are correspondingly reduced.  We therefore greatly improve on these earlier attempts concerning all-sky 3D catalogs based on the 2MASS sample, which results in generating the 2MASS Photometric Redshift catalog, as well as examining the possibility to go beyond the reach of this survey. Both are feasible thanks to the release of the data from the Wide-field Infrared Survey Explorer \citep[WISE,][]{WISE}, as well as increasing numbers of spectroscopic redshifts from such surveys as SDSS and the Galaxy And Mass Assembly \citep[GAMA,][]{GAMAabout}. \\

\section{Photometry and redshifts used}
\label{Sec: Photometry and redshifts}

\subsection{2MASS XSC}
The basic photometric dataset used in our study is the \textit{Two   Micron All Sky Survey Extended Source Catalog} (2MASS XSC; \citealt{2MASSXSC}). 2MASS \citep{2MASS} was a ground-based survey between 1997 and 2001, covering 99.998\% of the celestial sphere and delivering uniform, precise photometry and astrometry over the entire sky in the near-infrared $J$ (1.25~$\mu$m), $H$ (1.65~$\mu$m), and $K_s$ (2.16~$\mu$m) bandpasses. Observations were conducted from two dedicated 1.3~m diameter telescopes located at Mount Hopkins, Arizona, and Cerro Tololo, Chile. The 2MASS All-Sky Data Release includes 471 million source extractions in the Point Source Catalog, and over 1.6 million resolved objects in the Extended Source Catalog. Of the latter, more than 98\% are galaxies, and the remaining objects are mainly Galactic diffuse sources \citep{XSCz}. The 2MASS XSC was designed to satisfy the survey science requirements as outlined by \cite{2MASSXSC}, the most important being the reliability and completeness for unconfused regions of the sky. These were met for sources brighter than $K_s = 13.5$ ($\sim\!2.7$~mJy) and resolved diameters of galaxies above $\sim10$--$15''$. Until the advent of WISE, the 2MASS XSC was the largest all-sky astro- and photometric catalog of galaxies, and currently retains this status where extended-source photometry is concerned.

The complete 2MASS Extended Source Catalog contains 1,646,966 entries; however, a small fraction of these are artifacts, have erroneous or   null measurements. We have removed from our sample the sources that fulfill any of the following criteria: flag \ttt{vc=2} (non-extended Galactic sources, usually blends of stars, confirmed visually;   7383 objects); flag \ttt{cc\_flag=a} (artifacts, 122   sources); $J$, $H$ or $K_s$ band magnitude(s) \ttt{NULL} or excessively high (erroneous) -- over 170,000 objects, lacking   measurements mainly in the $H$ and $J$ bands. This gave us 1,471,442 sources for further analysis (the `2MASS good' sample). As the basic photometry we use the 20~mag arcsec$^{-2}$ isophotal fiducial elliptical aperture magnitudes (\ttt{j/h/k\_m\_k20fe}) together with relevant error estimation (1-sigma uncertainty for these magnitudes). The magnitudes were corrected for Galactic extinction with the use of the \cite{SFD} maps (SFD) and the $A_\lambda/E(B-V)$ coefficients taken from \cite{CCM89}.

Galactic extinction is generally not an important factor for the infrared window, however, the zone within $5^\circ$ of the Galactic Plane is susceptible to extinction errors and coefficient uncertainties, rendering potential systematics in the 2MASS photometry at these low Galactic latitudes.  We note that due to overestimation of the extinction by the SFD maps within $|b|<5^\circ$ (e.g.\ \citealt{Schroder07}), many of the sources therein are corrected improperly and they were further removed due to unreliable magnitudes, often reaching negative values. For those reasons, the strip of $|b|<5^\circ$, and $|b|<8^\circ$ or even wider   around the Bulge ($\ell>330^\circ$ and $\ell<30^\circ$), is usually masked out in all-sky cosmological applications of 2MASS data (cf.\ \citealt{Maller03,BCJM11}). In addition, issues are being raised regarding the SFD maps in general, for instance that they may overestimate the extinction over the entire sky, not only in the Galactic Plane \citep{SF11,Yuan13} and new dust maps are currently in development, in particular with the use of WISE data \citep{MF13}. This recalibration, once available, will however not be significant for the majority of 2MASS sources, but will be an important factor with the optical corrections.

Finally, an important remark is that for any cosmological analyses such as those mentioned in Section \ref{Sec: Introduction}, which require both uniformity and completeness over the entire sky, only a part of the total `2MASS-good' sample will be applicable. The completeness limit of 2MASS (in terms of extragalactic source counts) is $\sim\!13.9$ in the $K_s$ band \citep{XSCz} and galaxies beyond this magnitudes are usually removed whenever all-sky uniformity is needed \citep[e.g.][]{MBthesis}. There are approximately 1 million 2MASS-good galaxies with extinction-corrected magnitudes $K_s < 13.9$, covering 96\% of the sky, i.e.\ basically the entire celestial   sphere, except for the Galactic bulge area \citep[e.g.][]{GH12}. See   also the latter reference, as well as \cite{Maller03} and   \cite{XSCz}, for detailed plots of all-sky 2MASS XSC   distribution. The mean source density in the areas where the 2MASS   sample is complete amounts to $\sim26$ deg$^{-2}$.

\subsection{WISE}
\label{Subsec: WISE}
\begin{figure*}[!t]
\centering
\includegraphics[width=0.9\textwidth]{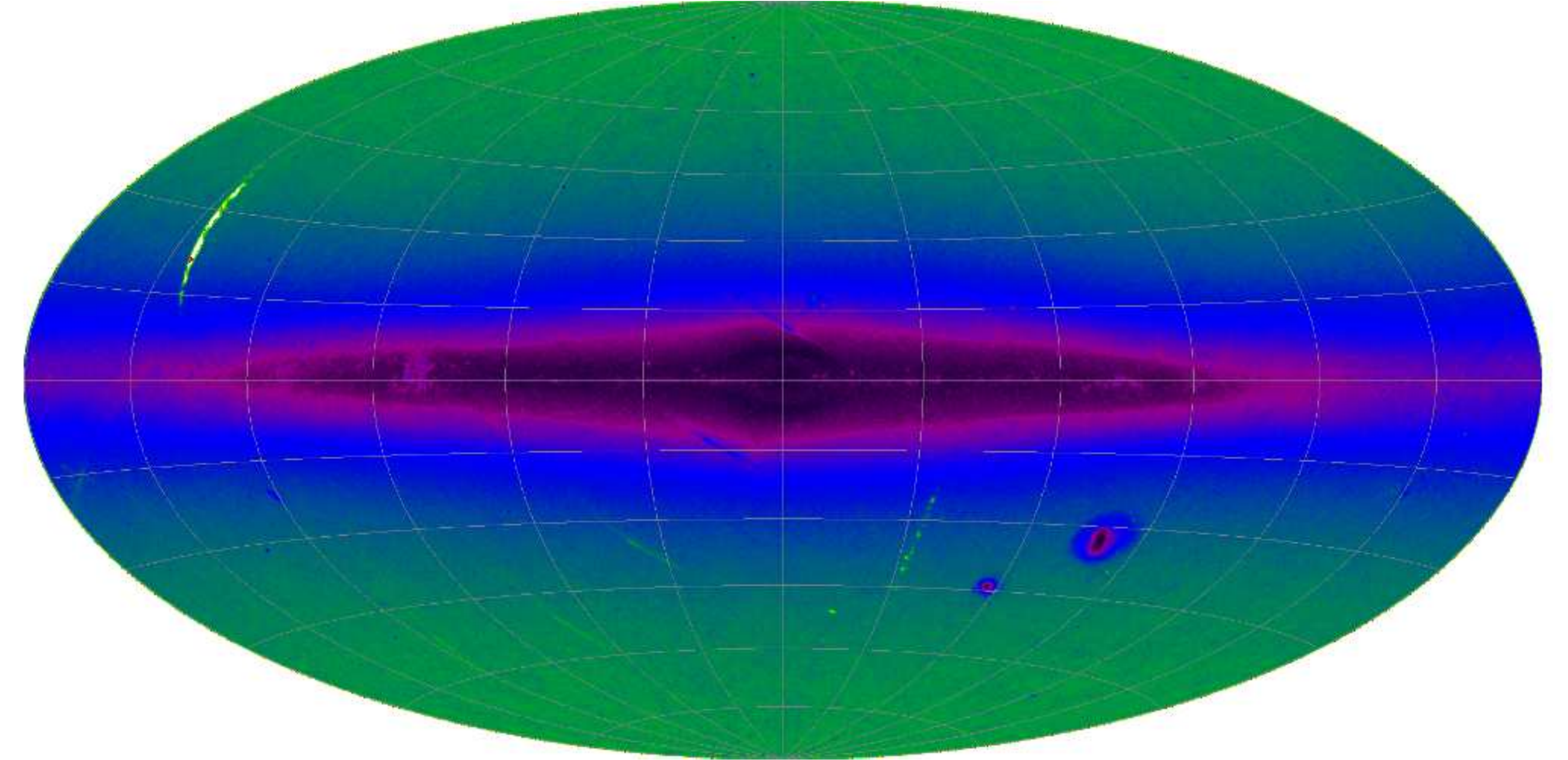}
\caption{\\Aitoff projection, in Galactic coordinates, of all-sky   source distribution in the WISE catalog, flux-limited at 15.5 in the   $W1$ band (more than 230 million objects in the plot). This figure   includes both extended and point-like sources.}
\label{Fig: WISE all-sky}
\end{figure*}

For the purposes of our project, the photometric data from 2MASS were supplemented by the measurements made within the WISE and SuperCOSMOS surveys. The former, the \textit{Wide-field Infrared Survey Explorer} \citep{WISE}, is a space-borne telescope that in 2010 mapped the entire sky at 3.4, 4.6, 12 and 22 $\mu$m ($W1$ -- $W4$). WISE achieved $5\sigma$ point source sensitivities better than 0.08, 0.11, 1 and 6 mJy in unconfused regions on the ecliptic in the four bands (but is considerably deeper at the ecliptic poles due to its polar orbit; see \citealt{Jarr11}), which is a major advance when compared to the now historical IRAS survey, as well as to the contemporary AKARI mission \citep{AKARI}, yet another orbital survey of the whole sky in the infrared. The all-sky WISE release \citep{WISEcat} includes all the data taken during the WISE full cryogenic mission phase that were processed with improved calibrations and reduction algorithms. The released data products contain a Source Catalog with positional and photometric information for over 563 million objects detected on the WISE images. A large fraction of these (well offset from the Galactic Plane) are galaxies, which makes WISE significantly deeper than 2MASS. The full potential of WISE extragalactic science remains to be explored: for instance, a `WISE XSC', which will incorporate aperture photometry of resolved (extended) sources, is now under construction \citep[see discussion in][]{Jarr13}. A particular example of early WISE XSC results is the WISE-GAMA catalog presented in Cluver et al.\ (2013, in prep.). 

Except for the case of the GAMA fields (see Subsection \ref{Subsec:   GAMA fields}), our basic photometric catalog is the 2MASS XSC. As we aim at as high a matching rate with WISE as possible, from the latter we use only the photometry in its two shortest wavelengths: $W1$ and $W2$ (although future updates my include $W3$ measurements as well). We expect that the galaxies present in 2MASS should always be seen in these shorter WISE bands, which have a much higher detection rate than the two other bands, $W3$ and $W4$ \citep{WISEexpl}. The overwhelming majority of all the objects in the WISE catalog have detections in the $W1$ band, and this rate is still fairly high for $W2$ (for $>2\sigma$ detections it is respectively 99.1\% and 83.5\% of the total). The matching rate of the `2MASS-good' sample with WISE $W1$- and $W2$-detected sources amounts in general to over 99\%. We pre-filtered the WISE catalog data by demanding that \ttt{cc\_flags[?]$\neq$`DPHO'} (no known artifacts; `\ttt{?}' stands for 1 or 2), \ttt{w?sat~$\leq 0.1$} (no more than 10\% of saturated pixels) and \ttt{w?snr~$\geq 5$} (sufficiently high profile-fit measurement signal-to-noise ratio -- equivalent to an upper limit on magnitude errors of $0.21$). Figure \ref{Fig: WISE   all-sky} presents an Aitoff Galactic projection of all WISE sources brighter than $W1=15.5$, pre-filtered as described above. There are over 230 million detections shown in this plot, of which 67 million are located at high Galactic latitudes of $|b|>20^\circ$. Note however that the WISE catalogs always contain a significant fraction of stars, even far away from the Galactic Plane \citep{Jarr11}, and especially at bright magnitudes. Apart from the Galactic Plane (with mostly stars), some other features clearly stand out: the Magellanic Clouds; lack of measurements in a strip with $\ell_\mathrm{Gal}\simeq 140^\circ$ and $15^\circ < b < 35^\circ$ due to `torque rod gashes'\footnote{\url{http://wise2.ipac.caltech.edu/docs/release/allsky/expsup/sec6\_2.html\#lowcoverage}} or some incompleteness due to moon contamination (see discussion in \citealt{Kovacs13}). We note that some of these biases are mitigated after matching WISE with the 2MASS XSC up to the completeness limit of the latter, as the resulting dataset is about $1.5$ magnitudes shallower than the one in Figure \ref{Fig:   WISE all-sky}. Additionally, the forthcoming post-cryo all-sky WISE release (`ALLWISE'), as well as a proper `WISE XSC', will alleviate at least some of these problems.

\begin{figure*}[!t]
\centering
\includegraphics[width=0.9\textwidth]{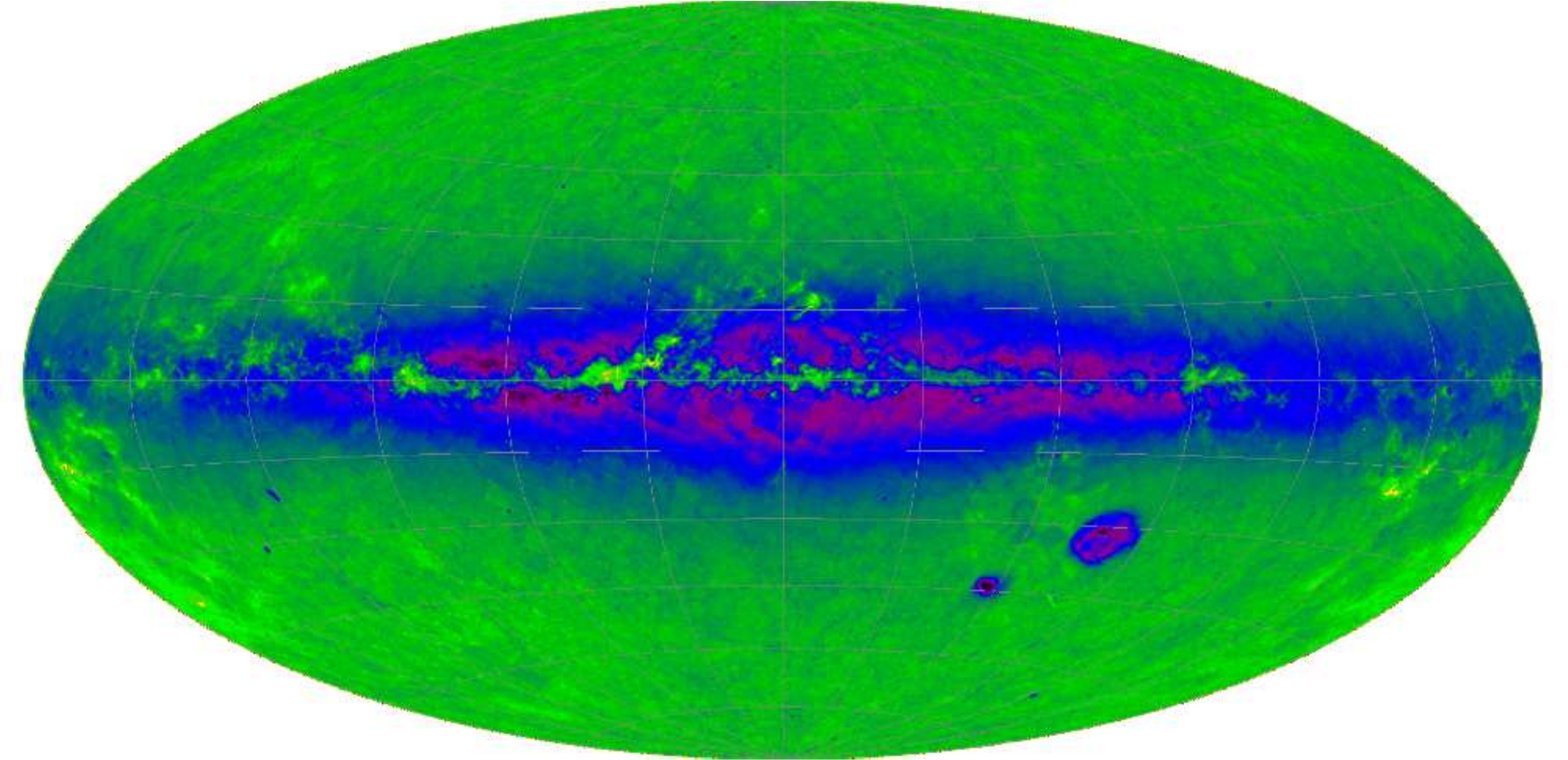}
\caption{\\Aitoff projection of all-sky extended source distribution   in the SuperCOSMOS survey, in Galactic coordinates. Objects below a   limiting magnitude of $B=21$ are shown (almost 189 million   detections). The spurious overdensities around the Milky Way and in   the Magellanic Clouds arise from blending.}
\label{Fig: SCOS all-sky}
\end{figure*}

The basic photometry in WISE is given by the \ttt{w?mpro} magnitudes measured with profile fitting appropriate for point sources. However, as emphasized in \cite{Jarr13} and in the Cautionary Notes of the WISE Explanatory Supplement\footnote{\url{http://wise2.ipac.caltech.edu/docs/release/allsky/expsup/sec1\_4.html}}, the Catalog contains both point-like and resolved sources, so the profile-fitting photometry underestimates the true brightness of extended objects. Our most recent findings show that probably all the 2MASS XSC galaxies are also resolved by WISE (Jarrett et al., in prep.)\ -- but extended-source photometry is currently available only for a fraction of thrm. Specifically, those WISE objects that are located within the 2MASS XSC perimeter -- defined to be the $K_s$-band isophotal circular diameter inflated by 10\% -- and have a radial distance between the WISE and 2MASS source of $<2''$, have integrated WISE fluxes measured in the elliptical aperture with a size, shape and orientation that is scaled from that of an associated 2MASS XSC source. These integrated magnitudes are given as \ttt{w?gmag} in the database and in principle could serve as a proxy for WISE galaxy fluxes before a proper extended source catalog is available for this survey. However, as only about half of all 2MASS-good sources have \ttt{w?gmag} measurements, we decided not to use this type of photometry in order to avoid biases in the photometric redshifts due to mixing extended- and point-source fluxes. Instead, until proper resolved source photometry is made available for the whole WISE catalog, we continue to use the \ttt{w?mpro} magnitudes, automatically correcting profile-fit photometry for extended emission using differential aperture photometry information from $5.5''$ and $11''$ circular aperture measurements, \ttt{w1mag\_1$-$w1mag\_3} \citep{WISEexpl}. This one additional parameter used as an input for the neural networks enables them to `learn' how to compensate for the flux lost due to the imposed assumption of a point source.

All the WISE W1 and W2 magnitudes were extinction-corrected using coefficients according to \cite{Flaherty07}: $A_{W1}/EBV = 0.231$ and $A_{W2} / EBV = 0.194$.

\subsection{SuperCOSMOS}
\label{Subsec:SCOS}

The \textit{SuperCOSMOS Sky Survey} (SSS) program \citep{SCOS1,SCOS2,SCOS3} was a project to digitize the photographic measurements of the sky in three band (optical $BRI$), via automatic scans of sky atlas photographs. The source photographic material came from the United Kingdom Schmidt Telescope (UKST) in the south and the Palomar Observatory Sky Survey-II (POSS-II) in the north, observations having been taken during the last quarter of the twentieth century. These catalogs were given an accurate photometric calibration with the use of SDSS photometry, overlaps of plates and by requiring uniformity in average color between the optical and 2MASS $J$ bands \citep{FP10}; the details of the procedure will be described in Peacock et al.\ (in prep.). Given suitable calibration information, these photographic plates are capable of yielding impressively accurate and uniform photometry. A good model for the rms residual between SuperCOSMOS photometry and SDSS predictions is obtained by adding in quadrature a constant value of $0.06$ and a magnitude-dependent $0.2\times   10^{0.4(m-m_{\rm lim})}$, where $m_{\rm lim}$ is the relevant $5\sigma$ completeness limit. The depth fluctuates between different plates, but typical figures for $m_{\rm lim}$ are approximately $B= 22$, $R= 21$, $I= 19.5$. 

The SuperCOSMOS magnitudes were given a zero point in which they would coincide with SDSS photometry in the case of an object with the colors of the fundamental SDSS standard. They are thus not a true AB system: see Peacock et al.\ (in prep.)\ for the color equations. As the passbands for the UKST and POSS-II were slightly different, in the magnitudes provided in the SuperCOSMOS database there is in effect a small color-dependent offset between the North and the South (i.e.\ above and below $\delta_{1950}=2.5^\circ$). By comparison with SDSS, we designed the following direct corrections to compensate for this effect and match the measurements in the two hemispheres, applied to the data from the South ($\delta_{1950}<2.5^\circ$):
\begin{equation}
B_S^{corr} = B - 0.03 (B - R)^2 + 0.03 (B - R) - 0.01 
\end{equation}
\begin{equation}
R_S^{corr} = R - 0.04 (B - R)^2 + 0.08 (B - R) - 0.02 
\end{equation}
\begin{equation}
I_S^{corr} = I - 0.02 (R - I)^2 - 0.07 (R - I) + 0.01
\end{equation}
where the subscript "$S$" refers to the south and right-hand sides include values taken directly from the catalog. Also the extinction-correction was hemisphere-dependent: we used $A^{N}_B=4.165$ and $A^{N}_R=2.773$ in the North, $A^{S}_B=4.011$ and $A^{S}_R=2.778$ in the South, and $A_I=2.016$ for the entire sky. In order to remove any residual biases not compensated for by this calibration, in the photometric redshift estimation process we trained separate neural networks for the two hemispheres.

The SuperCOSMOS data (SCOS for short) are curated at \url{http://surveys.roe.ac.uk/ssa/}, where multicolor information is provided for 1.9 billion sources, and extended source photometry is given whenever the object has been classified as such (\ttt{gCorMag}s in the database). We used only sources with \ttt{meanClass}~$= 1$ (source classified as a galaxy) and quality flags in the relevant bands below 2048 (no strong warnings, nor severe defects, \citealt{SCOS2}). In Figure \ref{Fig: SCOS all-sky} we plot the all-sky density map of SCOS extended sources, limited to observed $B<21$, which is one magnitude less than the estimated completeness limit. At these depths, there are almost 189 million sources with $B$ and $R$ extended-source photometry measured over the whole sky, although only those away from the Galactic Plane are applicable for extragalactic studies. For instance, there are over 50 million SCOS extended sources with $B<21$ at Galactic latitudes of $|b|>20^\circ$, which are mostly extragalactic. Here, unlike in the case of WISE, the increased density of detections near the Plane is due to spurious extended sources arising from blending of stars.\\

\subsection{Spectroscopic redshifts of 2MASS galaxies}
\label{Subsec: XSCspec}

About one third of all 2MASS galaxies already have spectroscopic redshifts, although their coverage on the sky remains far from uniform. In a study related to the discussion in this Subsection, \cite{LH11} presented the situation as it was in mid-2011, by compiling what they dubbed the \textit{2M++ catalog}, which included 69,160 2MASS-selected galaxies, reaching depths up to $K_s\leq12.5$ (although the effective depth varied on the sky). Their primary sources of redshifts were an early version of the 2MRS ($K_s\leq11.25$), the 6dFGS and the SDSS Data Release 7. Compiling the 3D data of the sources was only the starting point for \cite{LH11}, as they also assessed completeness in redshift for each of the regions and proposed the weights and corrections for incompleteness in redshift, sky coverage (Zone of Avoidance), as well as in apparent magnitude limits. Additionally, they presented the density field for their survey, and discussed the importance of large-scale structures such as the Shapley Concentration. In contrast, we prefer to recover individual photometric redshifts based on the \textit{observed} properties of our sources. This avoids the need for corrections such as `redshift-cloning' to mitigate small-scale redshift incompleteness arising from fiber collisions, as well as adjustment of apparent galaxy magnitudes due to the effects of cosmological surface brightness dimming and stellar evolution. The latter \textit{depend} on the redshift of the source and must not be applied if the photometric redshift technique is to work properly.
 
We compiled our primary spectroscopic redshift sample for 2MASS galaxies from the publicly available datasets described below. We used heliocentric redshifts throughout, because to work in the CMB frame it would be necessary to apply a Doppler $k$-correction to the photometry, which is redshift-dependent. Additionally, negative recession velocities in the heliocentric frame (although real for a tiny fraction of galaxies located within the Local Volume) are not of interest for our project and such sources are skipped.

The redshift surveys compiled within our project were the following:

\begin{itemize}

\item 2MASS Redshift Survey \citep[2MRS,][]{2MRS}, a uniform all-sky   (except for the ZoA) sample that contains 44,599 2MASS-selected   galaxies with $K_s\leq 11.75$ and $|b_\mathrm{Gal}| \geq   5^\circ$ ($\geq 8^\circ$ toward the Galactic bulge); of these,   43,361 are present in the `2MASS good' sample and have positive   heliocentric redshifts;

\item Sloan Digital Sky Survey Data Release 9 \citep[SDSS   DR9,][]{SDSS.DR9}, which includes 1,685,470 galaxy and quasar   redshifts, mainly in the North Galactic Cap   ($b_\mathrm{Gal}>20^\circ$), with some patches in the southern   Galactic hemisphere; there are 324,737 of these sources in the   `2MASS good' sample, with $z_\mathrm{hel}>0$;

\item 6dF Galaxy Survey Data Release 3 \citep[6dFGS DR3,][]{6dF.DR3},   a 2MASS-selected redshift and peculiar velocity survey that is   complete to total extrapolated magnitude limits $(K_s,H, J) =   (12.65, 12.95, 13.75)$ over 83\% of the southern sky, containing   125,071 redshifts in total; of these, 87,337 sources are   extragalactic, have adequate redshift quality and are present in the   `2MASS good' sample, with $z_\mathrm{hel}>0$;

\item 2dF Galaxy Redshift Survey \citep[2dFGRS,][]{2dF,2dFrelease},   with 226,906 galaxies of good-quality redshifts in an area of   approximately 1500 deg$^2$ selected from the extended APM   Galaxy Survey in three regions: a North Galactic Pole strip, a South   Galactic Pole strip and random fields scattered around the SGP   strip; of these, 46,150 are present in the `2MASS good' sample   within a matching radius of $3''$;

\item The ZCAT   compilation\footnote{\url{https://www.cfa.harvard.edu/\~dfabricant/huchra/zcat/}},   based on the CfA Redshift Catalog \citep{CfA}, updated throughout the   years by the late John Huchra. It contains 41,508 positive   heliocentric redshift from the whole sky, of which only 9,267 matched   2MASS-good sources within $3''$ (this low matching rate might be due   to inaccurate astrometry in the ZCAT).

\end{itemize}

\begin{figure}[b]
\centering
\includegraphics[width=0.45\textwidth]{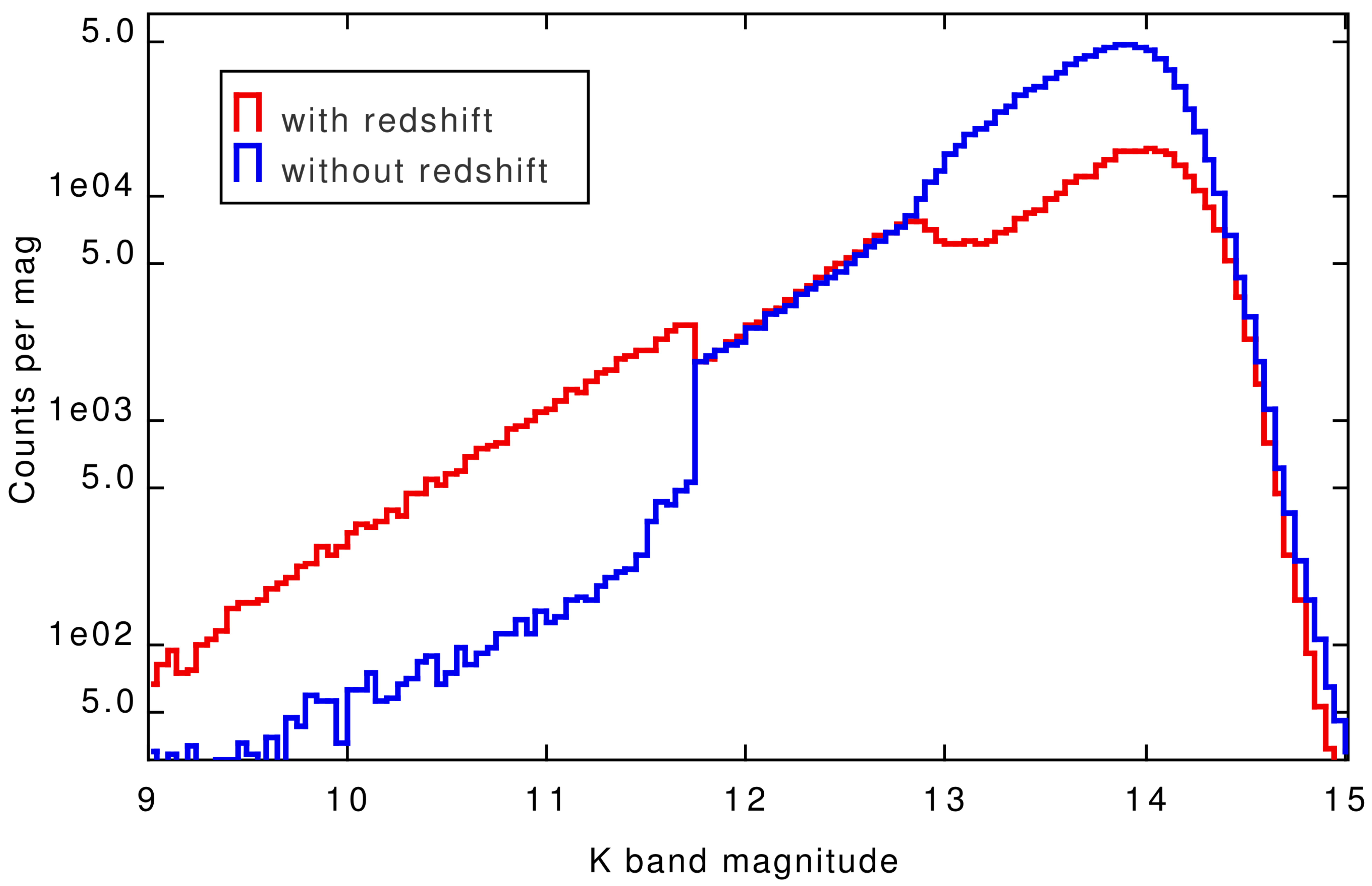}
\caption{Number counts as a function of the $K_s$ band magnitude for   those 2MASS XSC sources that have spectroscopic redshift   measurements (red line, 32\% of the total sample) and for those   that do not (blue line).}
\label{Fig: Mags hist}
\end{figure}

\begin{figure*}[!t]
\centering
\includegraphics[width=0.9\textwidth]{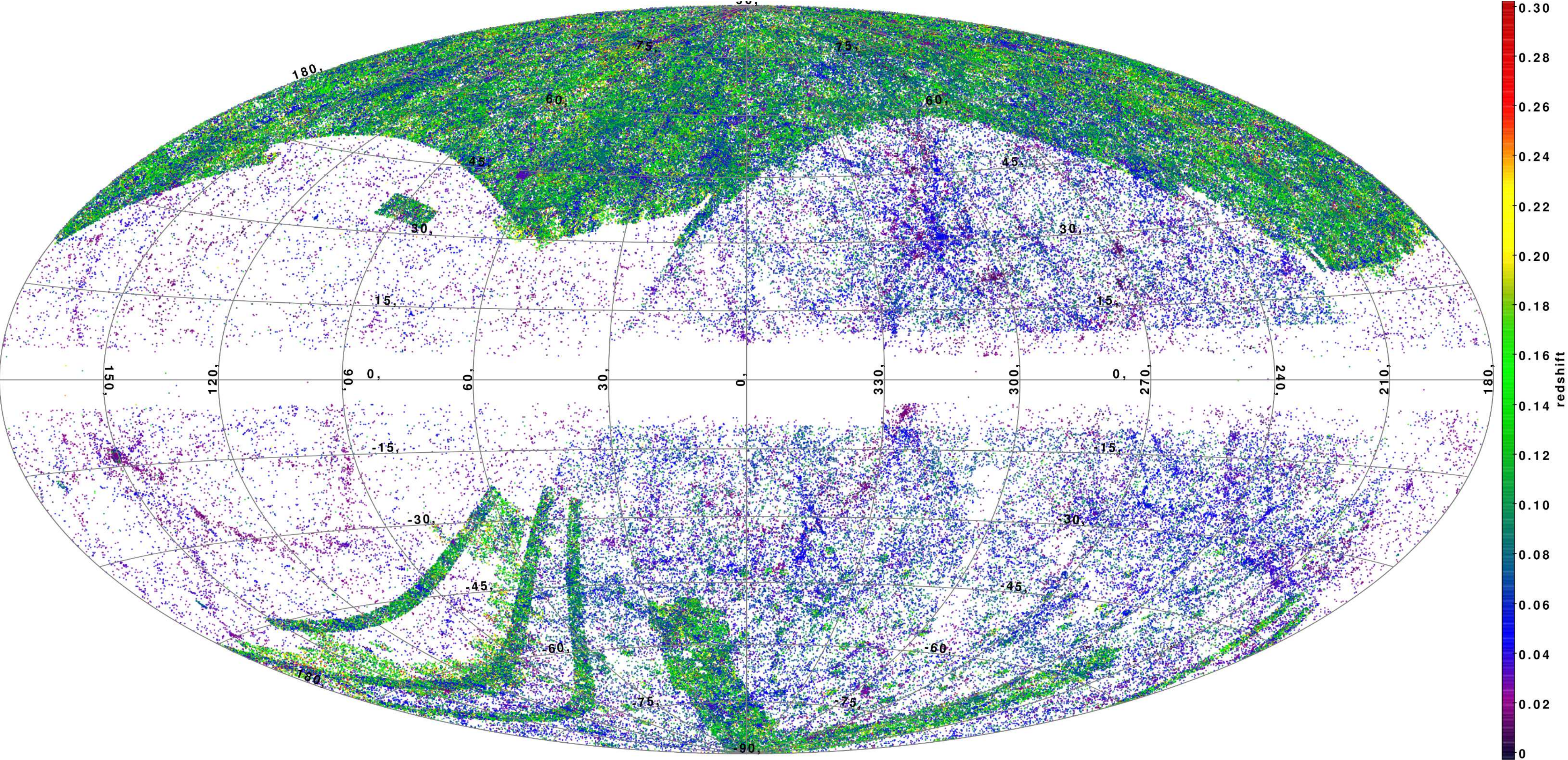}
\caption{\\Aitoff projection, in Galactic coordinates, of the all-sky   source distribution of the 2MASS-good galaxies for which   spectroscopic redshifts are presently available, as described in the   text. This plot shows almost 465,000 sources.}
\label{Fig: XSCspec all-sky}
\end{figure*}

These spectroscopic redshift surveys have some overlap, so that after adjusting for redundancy in total we obtained 464,857 positive heliocentric redshifts for the `2MASS good' sample, which makes 32\% of the total. In rare cases where conflicts in redshifts for the same object occurred, the priority was 2MRS $>$ SDSS $>$ 6dF $>$ 2dF $>$ ZCAT. The median redshift of the `2MASS-good-spectro' sample is $z_\mathrm{med}=0.082$.

In Figure \ref{Fig: Mags hist} we plot number counts as a function of the $K_s$-band magnitude for the 2MASS-good sources that have spectroscopic redshifts measured and for those that do not. This plot clearly shows that the spectroscopic data are not uniformly selected: the three `bumps' in the magnitude distribution of spectro-$z$ sources reflect the fact that the sample was compiled mainly from 2MRS, 6dFGS and SDSS. On the other hand, one can see that the spectroscopic redshifts are lacking only for objects with $K_s\gtrsim11.5$, thanks to the all-sky completeness of the 2MRS below these magnitudes. The fact that number counts in the magnitude slice $11.75<K_s<12.85$ are very similar for sources with and without spectroscopy is just a coincidence: approximately half of the 2MASS galaxies at these magnitudes have redshifts from 6dFGS and SDSS, while the other half is located in the north not covered by SDSS and in the ZoA.

Figure \ref{Fig: XSCspec all-sky} shows the Aitoff Galactic projection of all-sky distribution of the 2MASS-good sources with spectroscopic redshifts, color-coded by $z_\mathrm{spec}$. This plot illustrates the fact that redshift completeness occurs only in the areas sampled by SDSS. If an all-sky \textit{spectroscopic} redshift coverage of a depth comparable to 2MASS is ever to be obtained, a Southern analog of SDSS is essential, and the remaining gaps in the North need to be filled in. As things stand, the hope for progress towards this goal in the relatively near future relies on the success of the currently proposed TAIPAN survey \citep{CBB13}. What can be additionally seen from Fig.\ \ref{Fig: XSCspec all-sky} is that the large redshift surveys have been systematically avoiding the ZoA. Only the shallow 2MRS has probed into $|b|<10^\circ$, still excluding the Galactic equator and bulge areas. This means that we practically do not have calibration data for photometric redshifts at low Galactic latitudes: only $\sim$3\% of all the 2MASS sources in this strip have spectroscopic information. However, we are still able to calculate photo-$z$'s within the ZoA by training on other parts of the sky. Nevertheless, due to problematic photometry (caused by stellar crowding and overestimated extinction correction), the photometric redshifts in this area are often underestimated.

In view of deriving photometric redshifts for the entire 2MASS catalog, an important caveat is whether this combined spectroscopic redshift sample is representative of the whole underlying dataset. This is not to say that equal sampling at all redshifts is required, but the full range of redshifts must be spanned, and examples of all the galaxy types at a given redshift are required. The actual 2MASS redshift distribution is of course not known, but certainly some of the above-listed redshift surveys  will inevitably be shallower than an unbiased subsample of the full survey. Among the redshift datasets we use, the closest to the actual 2MASS redshift distribution will be the SDSS DR9 in the main (contiguous) area in the North. As we discuss in Subsection \ref{Subsec: NGalCap}, the majority (87\%) of all 2MASS galaxies located in this region have redshifts measured from that survey. In our future cosmological applications of the photometric redshift catalog presented here we will have to make a trade-off between representativeness of the spectroscopic redshift sample used for photo-$z$ estimation and its size. Note that there are, however, photometric redshift methods that, at least in principle, are independent of and do not require the usage of any training sets, such as for instance the EAZY package \citep{EAZY}. 

In addition to the large-area redshift surveys listed above, which give the basis for the all-sky 2MASS photometric redshift estimation, we additionally used the recent data from the GAMA Data Release 2 (DR2; Liske et al.\ 2013, in prep.)\ to investigate the possibility of going deeper than with 2MASS XSC, while retaining all-sky coverage. This GAMA data release provides AAT/AAOmega spectra, redshifts and a wealth of ancillary information for 72,225 objects from the first phase of the GAMA survey (2008 -- 2010, usually referred to as GAMA~I). The GAMA~I survey extends over three equatorial survey regions of 48 deg$^2$ each, called G09, G12 and G15 \citep{GAMAinput}. In DR2, data for all GAMA~I main survey objects with $r < 19.0$ (G09 and G12) or $r < 19.4$ (G15) were released. We used the 70,330 DR2 sources with positive redshifts of quality $NQ>2$. The median redshift of this sample is $z=0.171$. Although it covers just a small fraction of the sky, its depth and completeness -- better than that of the SDSS main galaxy sample -- make GAMA an ideal catalog for all-sky photometric redshift calibration at larger distances than possible with 2MASS.

\section{Photometric redshift framework}
\label{Sec: Photo-z framework}

Methods for estimating photometric redshifts are now abundant, and fall under two general categories (and combinations thereof): machine learning and spectral energy distribution (SED), or template, fitting. The first approach, proposed originally by \cite{Connolly95}, derives an empirical relation between magnitudes (and/or possibly other galaxy parameters, such as colors) and redshifts using a subsample of objects with measured spectroscopic redshifts, i.e.\ the \textit{training set}. The trained algorithm (often additionally validated on a separate \textit{validation set}) is then applied to the rest of the sample for which no spectroscopic redshifts are known. Machine learning methods generally rely on the availability of a sufficiently large and representative training set in order to be efficient and unbiased. However, they often do not require many photometric bands to perform successfully and even adding noisy measurements usually improves their performance.

The second general approach, the SED fitting procedure \citep[e.g.][]{BMP00}, bases its efficiency on the fit of the overall shape of spectra and on the detection of strong spectral features. The observed photometric SEDs are compared to those obtained from a set of reference spectra, using the same photometric system. The photometric redshift of a given object corresponds to the best fit of its photometric SED by the set of template spectra. This method may be advantageous over machine learning as it usually does not require training sets. On the other hand, it needs accurate photometry -- especially matching of seeing and apertures between passbands -- and some prior knowledge of expected redshift distribution to perform efficiently.

As observed by \cite{FP10}, photometric redshift estimation methods usually yield $z_\mathrm{phot}$ that are unbiased in a sense that the mean true redshift at given $z_\mathrm{phot}$ is equal to $z_\mathrm{phot}$. But in order for this to be true, there must then be a bias in $z_\mathrm{phot}$ at given true $z_\mathrm{spec}$, unless redshift distribution is constant. Since $N(z)$ in practice has a well-defined maximum, this means that $z_\mathrm{phot}$ will be overestimated near $z= 0$ and underestimated at high $z$.

Some of the particular implementations of the methods listed above have been compared in \cite{ABLR11}, who applied them to $\sim\!1.5$ million luminous red galaxies in SDSS DR6. Partially motivated by their results and by the fact that a significant fraction of our dataset already has spectroscopic redshifts measured, we employ the machine-learning approach. We have also made some early attempts into extending our study to SED-fitting methods, by testing the EAZY package \citep{EAZY} on our data. Preliminary results are however not competitive with those described below, with accuracies typically 50\% worse than in the machine-learning case. We suspect that this would arise also for other template-matching methods if used with our samples. The synthetic photometry may not match the real one for various reasons: the template spectra could be in error; we do not know the filter transmission curves perfectly (including the wavelength-dependent photographic sensitivity and the contribution of atmospheric absorption); aperture corrections may be slightly different in different bands if the seeing is wavelength-dependent. \ANNz\ by its nature of using training sets from the data itself effectively self-calibrates and mitigates such systematics.

The need for SED templates was avoided by \cite{FP10}, who took a highly direct approach to the generation of photometric redshifts, averaging over neighbors of known redshift at a given location in the five-band magnitude space. By using magnitudes rather than colors, this automatically built in information from the luminosity function, so that bright galaxies were never allocated an extreme redshift that would require them to have an unrealistic luminosity. As a result, the scatter in photometric redshift declined towards $z=0$. The overall rms in $z_\mathrm{phot}$ -- $z_\mathrm{spec}$ in the \cite{FP10} analysis was $0.033$. This method is however limited by the need to bin the data, which requires an extremely large training set.

A non-exhaustive list of alternative photometric redshift estimators that use training sets includes such methods as Artificial Neural Networks \citep[ANN,][]{FLS03,Tagliaferri03,ANNz,Vanzella04,Way09,Cavuoti12}, Support Vector Machines \citep{Wadadekar05}, Virtual Sensors \citep{WS06}, Random Forests \citep{Carliles10}, Boosted Decision Trees \citep{Gerdes10}, Weak Gated Experts \citep{Laurino11} or Self-Organized Maps \citep{Geach12,WK12}. Throughout our analysis, we applied a particular implementation of the first of these, namely the \ANNz\ \citep{ANNz}, leaving a possible comparison of different algorithm performance in our particular application for future work.

{The \ANNz\ package is freely available, ready-to-use software\footnote{\url{http://www.homepages.ucl.ac.uk/\~ucapola/annz.html}} for photometric redshift estimation using artificial neural networks.  For a detailed description of the \ANNz\ algorithm and underlying methods, see \cite{ANNz} and references therein. In short, it learns the relation between photometry and redshifts from appropriate training sets of galaxies for which the spectroscopic redshifts are already known. Whenever a sufficiently large} and representative training set is available, \ANNz\ is a highly competitive tool when compared with traditional template-fitting methods \citep{FLS03,ANNz,WHG09,ABLR11} and has been used for various data sets, including SDSS \citep{Collister07}, GAMA \citep{Christo12} or the Blanco Cosmology Survey \citep{Desai12}. As we show in the remaining analysis, also in our application the \ANNz\ does a very commendable job as gauged in both systematic and random errors.

\section{Photometric redshift results: test fields for 2MASS and beyond}
\label{Sec: Photo-z tests}

Before constructing the 2MASS Photometric Redshift catalog, which will be described in Section \ref{Sec: 2MPZ}, we selected three test fields to verify how photometric redshift estimation performs for 2MASS-based data. We aimed at choosing such areas on the sky that would contain a sufficiently high number of galaxies each to clearly demonstrate the potential of the machine-learning methods for all-sky photometric redshift estimation. Additionally, these fields were defined in a way to emphasize possible biases and caveats, such as non-representativeness of the underlying spectroscopic redshift sample for the whole photometric set or the proximity of the Galactic Plane (ZoA) and resulting high extinction or stellar contamination. Finally, separate treatment of the GAMA fields served as a test-bed for our future work that will be based on WISE instead of 2MASS XSC data. The test fields discussed below are:

\begin{itemize}
\item North Galactic Cap (NGCap), defined as a circular region with $b_\mathrm{Gal}> 60^\circ$;
\item South Galactic Cap (SGCap), a circular region with $b_\mathrm{Gal}< -60^\circ$;
\item North Ecliptic Cap, a circular region for which $b_\mathrm{ecliptic}> 60^\circ$;
\item GAMA equatorial regions, composed of 3 rectangles enclosed within the following equatorial coordinates:  (G09) $129^\circ\leq \alpha \leq 141^\circ$ and $-1^\circ \leq \delta \leq 3^\circ$; (G12) $174^\circ\leq \alpha \leq 186^\circ$ and $-2^\circ \leq \delta \leq 2^\circ$; (G15) $211^\circ\leq \alpha \leq 224^\circ$ and $-2^\circ \leq \delta \leq 2^\circ$.
\end{itemize}

As the purpose of the present Section is to verify the performance of \ANNz\ in view of creating all-sky photometric redshift samples, the 'spectroscopic-only' sample for each field was further divided (randomly) into disjoint training, validation and test sets, respectively proportioned as 5:1:12. The proportion 1:2 between the calibration (training+validation) and test samples was chosen to mimic the actual situation with the 2MASS data, where every third galaxy has the spectroscopic redshift measured. Relative sizes of the training and validation sets were chosen somewhat arbitrarily, which however has no significant influence on the results, as long as both are sufficiently large, which is always the case here. The test sets were not used in the photo-$z$ calibration procedure and served only for comparison purposes. 

In addition to the preliminary selections applied to 2MASS, WISE and SCOS data described in Section \ref{Sec:   Photometry and redshifts}, we additionally cleaned the samples used for the photometric redshift calibration  by removing the sources fulfilling any of the following conditions: $z_\mathrm{spec}\leq0.003$ and high-redshift outliers; errors in any of the magnitudes $\Delta(mag)\geq0.2$; relative errors in $z_\mathrm{spec}$ (when available) larger than 10\%; reddening $E(B-V)\geq 0.25$. This filtering removed however only a small fraction of the spectroscopic samples.

The optimal architectures of the neural networks depended on the field and varied with number of inputs and size of training sets, but usually two intermediate layers with 6 to 23 nodes each sufficed. Committees of 6 to 30 networks, each trained with a different initial seed, were used. The inputs for the \ANNz\ were the magnitudes in the photometric bands used, i.e.\ $J,H,K_s$ from 2MASS, $W1,W2$ from WISE and $B,R,I$ from SCOS, supplemented with WISE differential photometry \ttt{w1mag\_1$-$w1mag\_3} as an additional parameter whenever WISE data were used. The latter input allowed to automatically correct WISE profile-fit photometry for extended emission, see the discussion in Sec.\ \ref{Subsec: WISE}. We checked that employing for instance a 2MASS angular size in addition to the WISE differential aperture magnitude gives tiny gain in photo-$z$ accuracy, while every new parameter for the training considerably inflates computation time, especially for the full-sky sample.

{We have also verified whether additional input parameters, uncorrelated with the spectroscopic redshift, could help mitigate some possible systematics. The latter were identified by a posteriori checks of the photometric redshift samples; in particular, at low Galactic latitudes, wherever both extinction is severe and local star density is high, a small fraction of photo-$z$'s are significantly underestimated. Looking for possible remedies, we considered the following inputs for the \ANNz: Galactic extinction, angular coordinates and local star density\footnote{2MASS coadd log(density) of stars with $K<14$.}. None of them helped remove the bias nor improve overall statistics. The reason is that the photometry of the discussed sources is systematically compromised: first, the \cite{SFD} maps are known to overestimate extinction corrections at low Galactic  latitudes; second, galaxy fluxes are polluted by Milky Way stars often superimposed on galaxy images. Both these effects cause the problematic sources to be seen as bluer and brighter than they actually are. This results in their photometric redshifts being systematically too low, which cannot be overcome by `regularizing' \ANNz\ performance with additional input parameters.}

Each test field, as well as the all-sky sample, was treated separately, in a sense that the neural networks were optimized to each specific case by being trained independently. In addition, as discussed in Subsec.\ \ref{Subsec:SCOS}, separate networks were trained for the sources above and below $\delta_{1950}=2.5^\circ$.

In order to quantify how \ANNz\ performs, we calculate and report the following statistics for the test samples in each of the regions:
\begin{itemize}
\item mean $\langle z \rangle$  and median $\overline{z}$ redshift of the true $z_\mathrm{spec}$ (input) and the calculated $z_\mathrm{phot}$ (output) distribution;
\item 1$\sigma$ scatter\footnote{Note that we do \textit{not} remove outliers in $\delta z$ prior to calculating this statistic, so it may overestimate the actual scatter.} between the spectroscopic and photometric redshifts, $\sigma_{\delta z}=\langle (z_\mathrm{phot}-z_\mathrm{spec})^2 \rangle ^{1/2}$;
\item scaled median absolute deviation, defined as  $\mathrm{SMAD}(\delta z) = 1.4826 \times \mathrm{med}(|\delta z-\mathrm{med}(\delta z)|)$, a more robust measure of statistical dispersion, less prone to outliers than  $\sigma_{\delta z}$;
\item net bias of $z_\mathrm{phot}$: $\langle \delta z \rangle = \langle z_\mathrm{phot}-z_\mathrm{spec} \rangle$;
\item median of the relative  error, where the latter is defined as $\Delta z = |z_\mathrm{phot}-z_\mathrm{spec}| / z_\mathrm{spec}$ (expressed in per cent);
\item percentage of outliers for which $|z_\mathrm{phot}-z_\mathrm{spec}| > 3\, \mathrm{SMAD}({\delta z})$.
\end{itemize}

{Note that we have chosen to use the term 'outliers', without identifying them as 'catastrophic', as the latter are ambiguously defined in the literature (see e.g.\ \citealt{Ilbert06,EAZY,Oyaizu08,Brescia13} for different variants). If we chose to count the sources with $|z_\mathrm{phot}-z_\mathrm{spec}| > 3\, \sigma_{\delta z}$ instead of those outside $3\, \mathrm{SMAD}({\delta z})$, as in \cite{Oyaizu08} and \cite{Gerdes10}, the numbers quoted below would decrease 2--3 times to typically 1\%.}

For the first three fields (the Caps), we present results of the photo-$z$ estimation based on from 5 to 8 bands in total plus one WISE differential photometry parameter where relevant. Including WISE and especially SCOS in addition to 2MASS greatly improves the statistics of the $z_\mathrm{spec}$ -- $z_{\mathrm{ANN}z}$ comparison. On the other hand, adding SCOS slightly lowers the matching rate since not all the 2MASS\ti{}WISE sources are also present in SCOS with reliable photometry.  

For the GAMA regions, however, we experiment with matching only WISE with SCOS, thus attempting measurements at greater depths (further described below, Subsection \ref{Subsec: GAMA fields}).  This exploratory work gives a first taste of what will be possible when WISE is matched with higher-fidelity samples such as Pan-STARRS \citep{Pan-STARRS}, SkyMapper \citep{SkyMapper} or VISTA-VHS \citep{VHS}.

\begin{figure*}[t]
\centering
\subfigure[]
{
\includegraphics[width=0.319\textwidth]{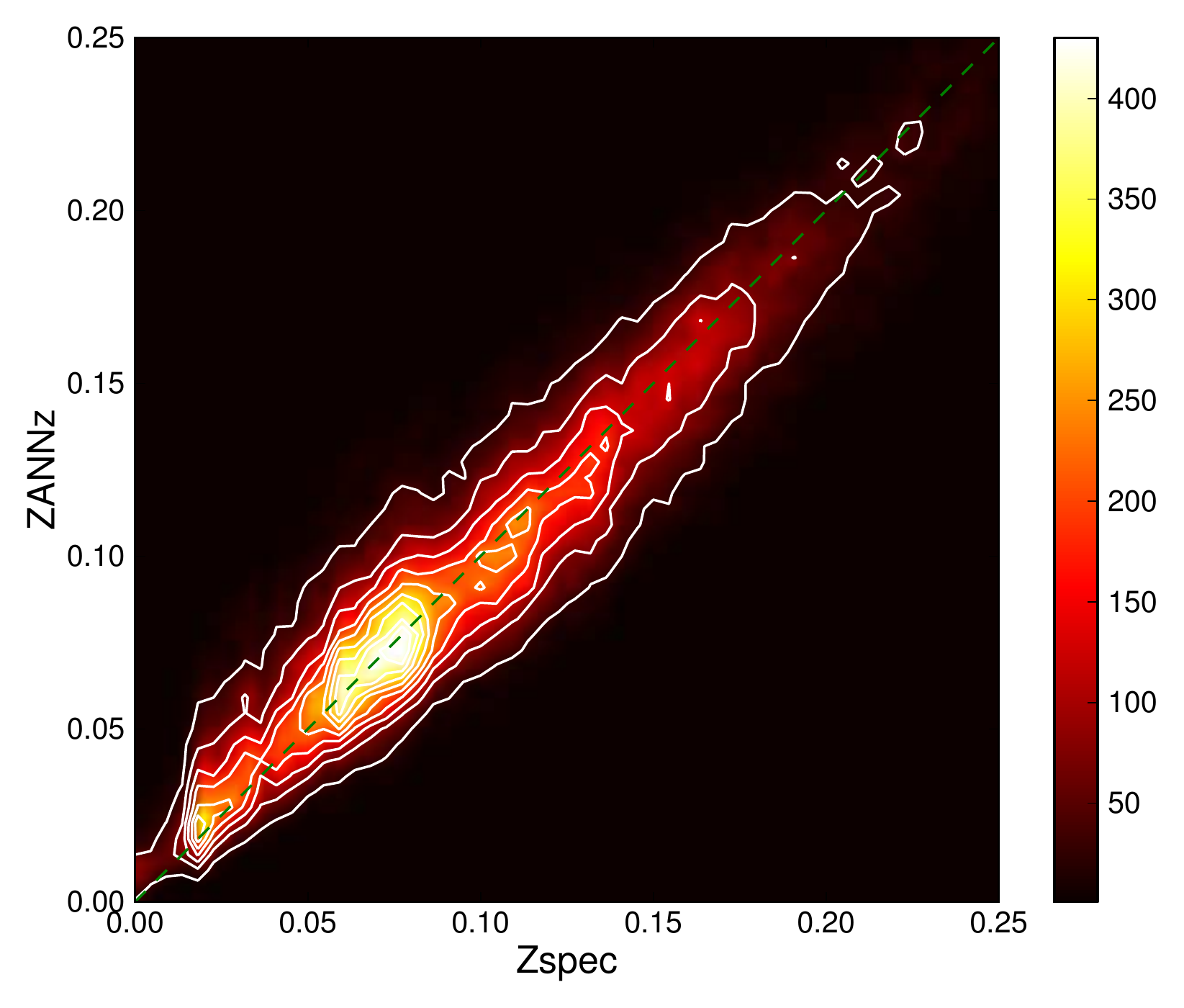}
\label{Fig: NGalCap 2MxWIxSC density}
}
\subfigure[]
{
\includegraphics[width=0.319\textwidth]{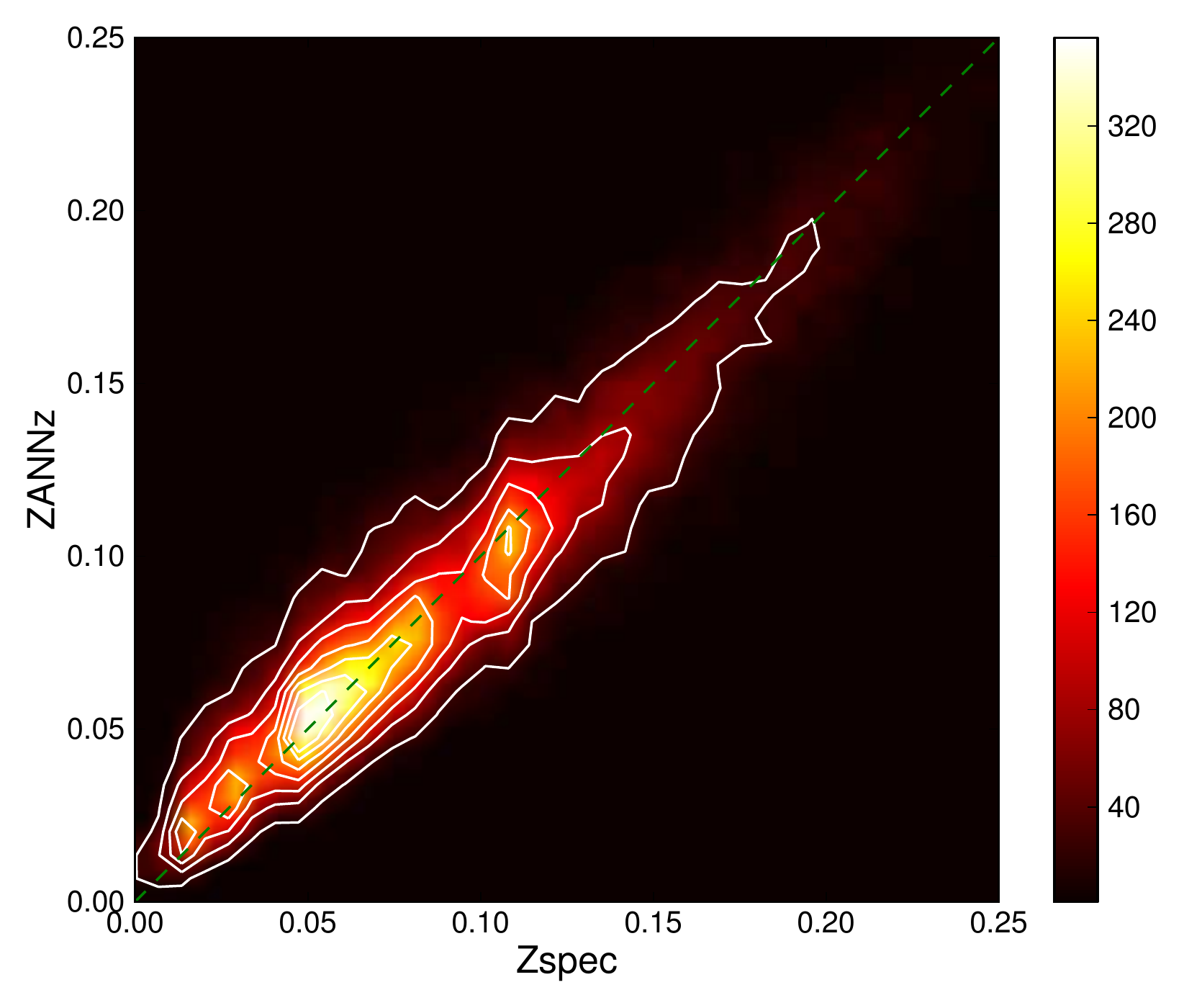}
\label{Fig: SGalCap 2MxWIxSC density}
}
\subfigure[]
{
\includegraphics[width=0.319\textwidth]{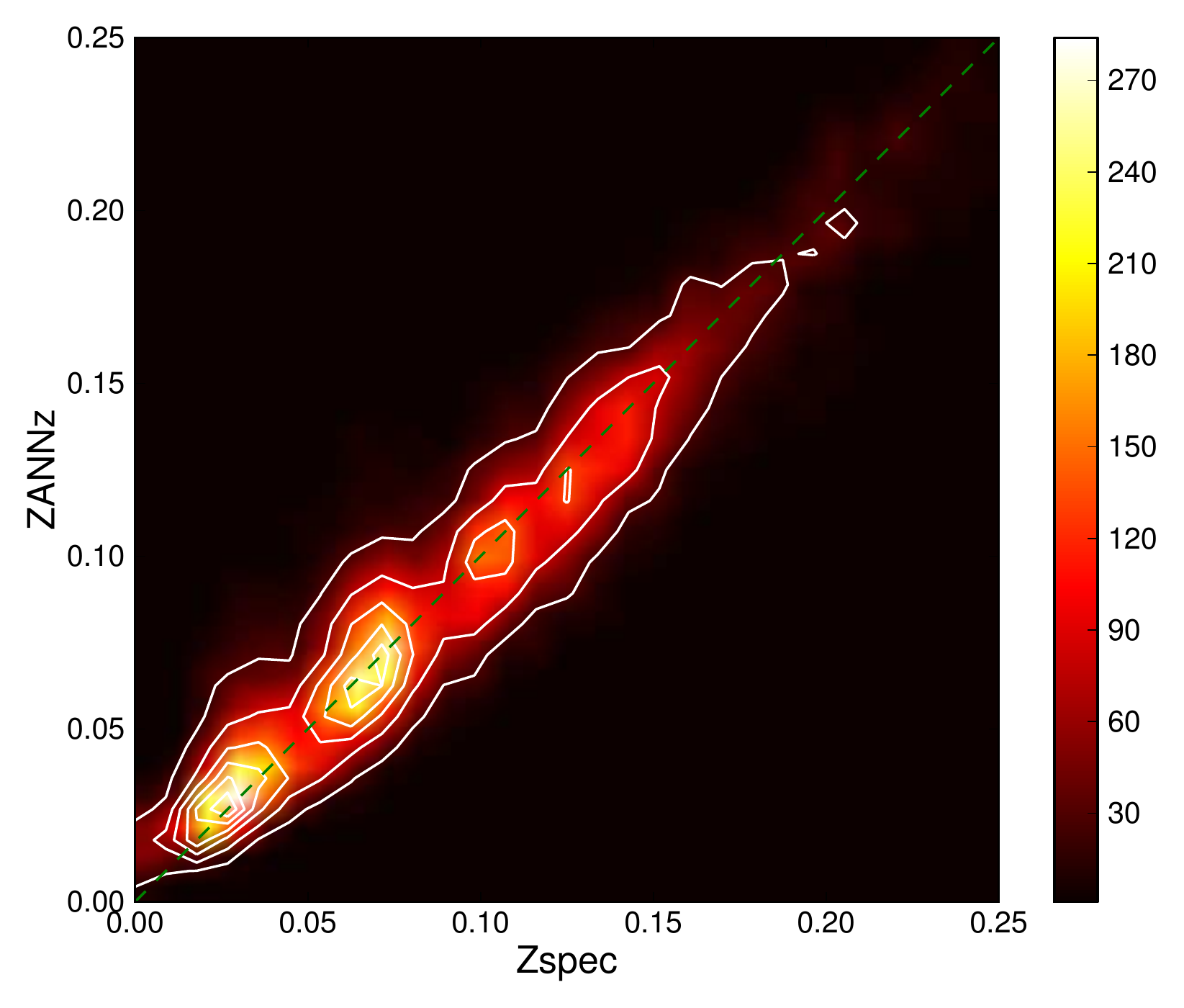}
\label{Fig: NEclCap 2MxWIxSC density}
}
\caption{Comparison of spectroscopic and photometric redshifts for   2MASS\ti{}WISE\ti{}SCOS matches (8 photometric bands) in: (a) the   North Galactic Cap, $b_\mathrm{Gal}>60^\circ$ (over 74,000 sources   in the test set); (b) the South Galactic Cap,   $b_\mathrm{Gal}<-60^\circ$ (over 26,000 sources in the test set);   the North Ecliptic Cap, $b_\mathrm{ecl}>60^\circ$ (roughly 9,000   sources in the test set).}
\end{figure*}

\subsection{North Galactic Cap}
\label{Subsec: NGalCap}

This region of $\sim\!2800$~deg$^2$ contains 131,707 `2MASS-good' sources.  Of these, over 99\% have matches with the 'WISE clean' sample within a $3''$ radius and more than 97\% of the total were identified in the SCOS as having all three $B R I$ extended-source magnitudes measured. The combined 2M\ti{}WI\ti{}SC data set includes more than 127k of the original 2MASS XSC sources.

Our NGCap is an area fully covered by the SDSS main (contiguous) region and the majority of the 2MASS objects (87\%) have spectroscopic redshifts measured from this survey, plus some from other catalogs. Note however that in the eventual all-sky applications these samples will be smaller, as the 2MASS catalog needs to be flux-limited at $K_s \lesssim 14$ to preserve its uniform depth on the sky (the full dataset is deeper in the North, for $\delta \gtrsim 12^\circ$). Here, we do \textit{not} apply any flux limit as it has no important influence on our tests other than lowering the number of sources used for the photo-$z$ estimation.

\begin{figure*}[t]
\centering
\subfigure[]
{
\includegraphics[width=0.4\textwidth]{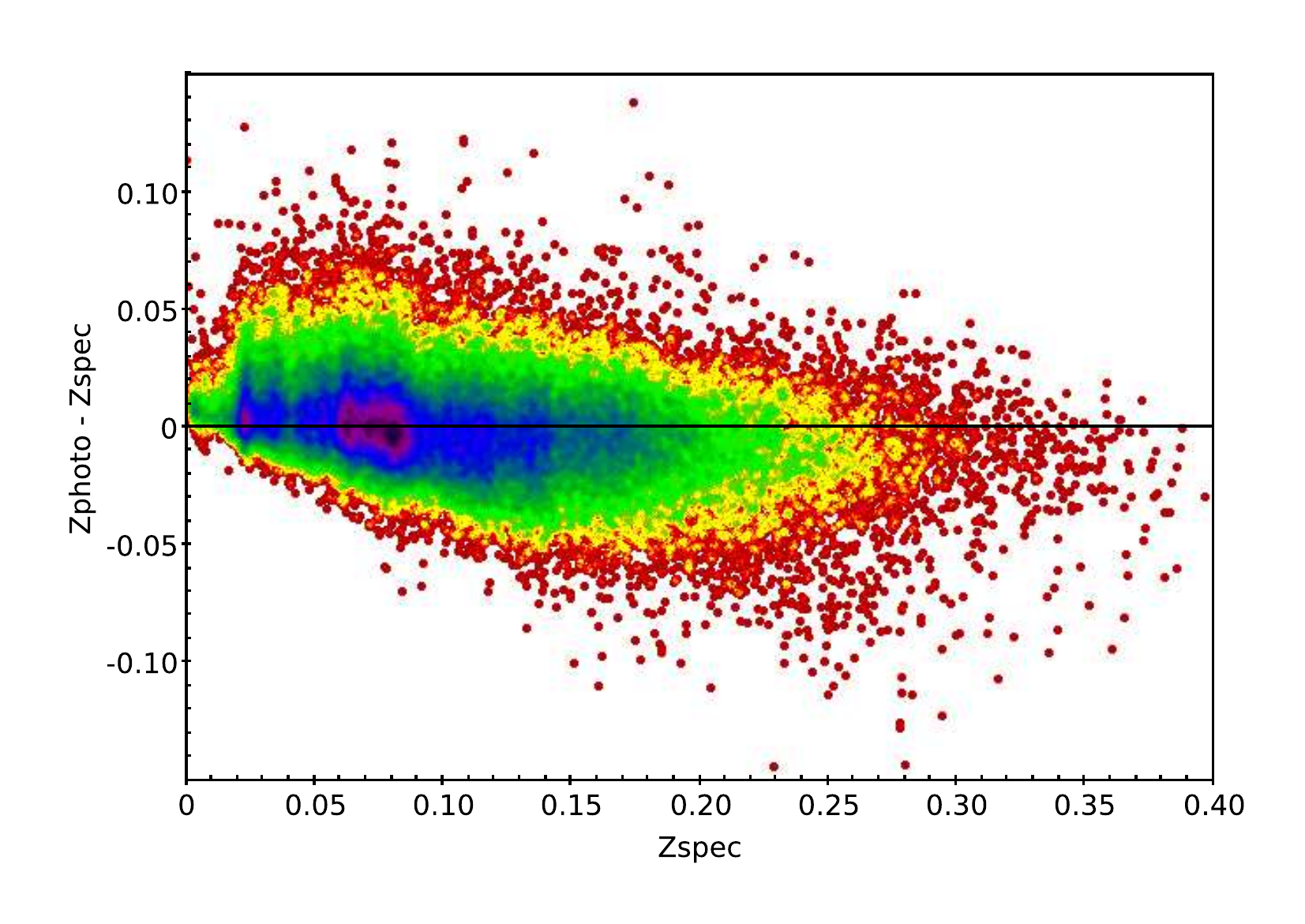}
\label{Fig: NGalCap zsp.biases}
}
\subfigure[]
{
\includegraphics[width=0.4\textwidth]{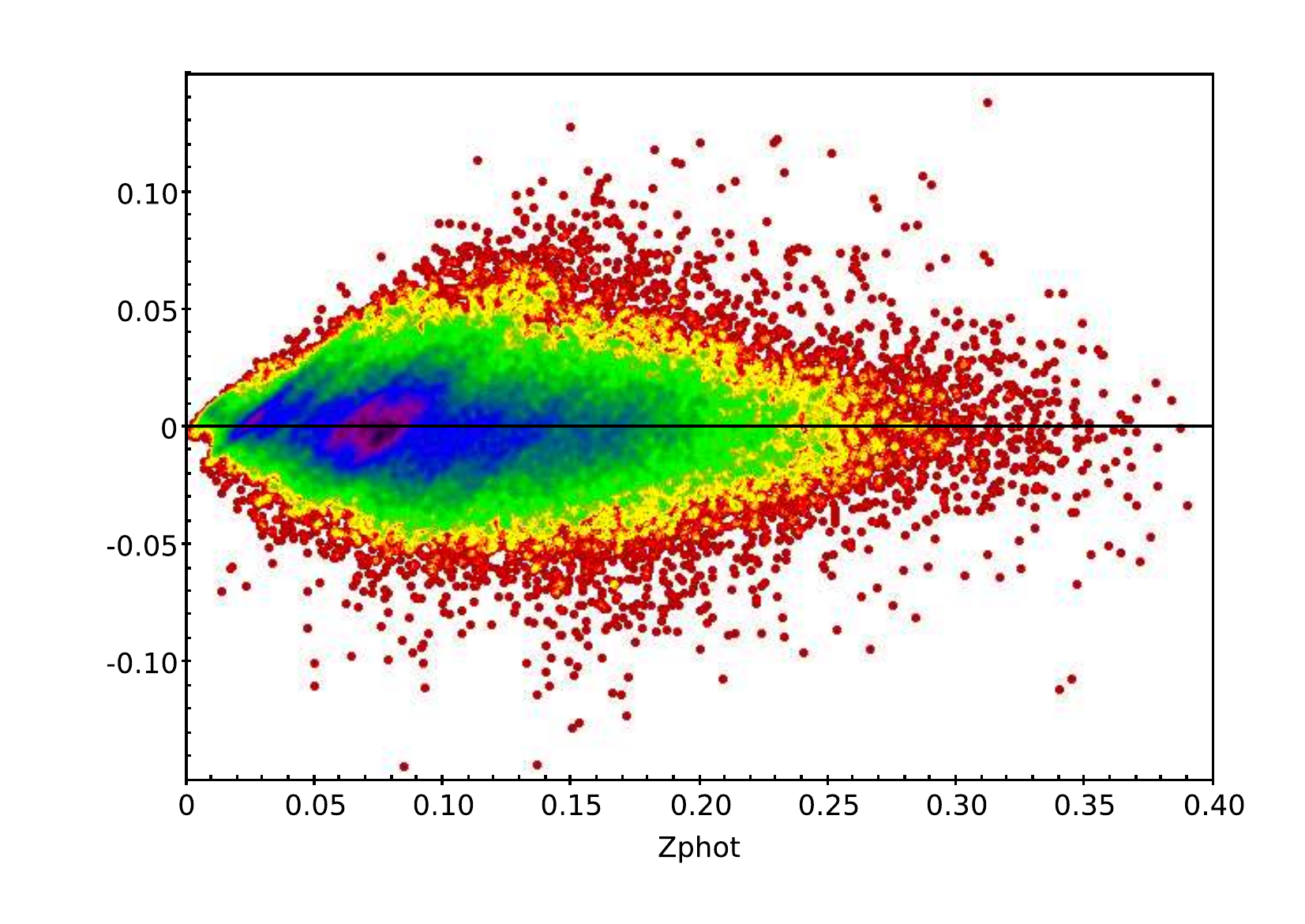}
\label{Fig: NGalCap zph.biases}
}
\caption{Error in the photo-$z$ as a function of (a) spectroscopic redshift and (b) photometric redshift. These plots illustrate the general property of the photo-$z$'s being unbiased in a sense   that the mean true redshift at given $z_\mathrm{phot}$ is equal to   $z_\mathrm{phot}$, while they are overestimated near $z_\mathrm{spec}= 0$ and underestimated at   high $z$.}
\end{figure*}

Having applied all the selection on the data, for the 2M\ti{}WI\ti{}SC sample we were left with fewer than 31k NGCap sources for neural network training and about 74.5k objects in the test set, and slightly larger respective sets for the 2M\ti{}WI and 2M\ti{}SC matches. The statistics for the \ANNz\ application are summarized in Table \ref{Table 2MASS}. We can note that: (a) already the 2MASS\ti{}WISE sample allows for a photo-$z$ accuracy comparable to that achieved by \cite{FP10} from 2MASS\ti{}SCOS matching (using only $B$ and $R$ bands); (b) the 2MASS\ti{}SCOS($BRI$) option gives better results, at a cost of a slightly lower matching rate; and (c) combining the three surveys together greatly improves the photo-$z$ estimates, which makes it the optimal approach. In particular, here as well as in the other test fields, supplementing 2MASS\ti{}WISE with the three SCOS bands ameliorates overall photometric redshift precision by 30\% or better.

The results are visualized in the density plot (Figure \ref{Fig: NGalCap 2MxWIxSC density}), which compares spectroscopic and photometric redshifts for the NGCap region. Since our photo-$z$ estimation uses magnitudes and not colors, the same effects as discussed in \cite{FP10} are observed. The scatter in photometric redshifts declines towards $z=0$ and the mean true (spectroscopic) redshift at a given $z_\mathrm{phot}$ is equal to $z_\mathrm{phot}$; but the photometric redshifts are overestimated near $z= 0$ and underestimated at high $z$. This is illustrated in Figs.\ \ref{Fig: NGalCap zsp.biases} and \ref{Fig: NGalCap zph.biases} and applies equally to all the datasets discussed below.

\begin{deluxetable*}{lcccccccccc}
\tabletypesize{\footnotesize}
\tablewidth{0pt}
 \tablecolumns{12} 
\tablecaption{\label{Table 2MASS}\small Statistics for the photometric redshift estimation for the 2MASS-based test samples.}

\tablehead{ 
\colhead{cross-matched} & 
\colhead{number of} & 
 \multicolumn{2}{c}  {mean $\langle z \rangle$} & 
 \multicolumn{2}{c} { median $\overline{z}$} &
 \colhead{ 1$\sigma$ scatter\tablenotemark{e}} & 
 \colhead{ scaled } &
 \colhead{ net bias\tablenotemark{g} } & 
 \colhead{ median } &
 \colhead { \% of } \\
\colhead{surveys\tablenotemark{a}} &
\colhead{sources\tablenotemark{b}} &
\colhead{spec\tablenotemark{c}} & \colhead{phot\tablenotemark{d}} &
\colhead{spec\tablenotemark{c}} & \colhead{phot\tablenotemark{d}} &
\colhead{ $\sigma_{\delta z}$} &
\colhead{ MAD\tablenotemark{f} } &
\colhead{$\langle \delta z \rangle$} &
\colhead{ error\tablenotemark{h} } &
\colhead{ outliers\tablenotemark{i} }
}

\startdata

\sidehead{\bf North Galactic Cap}
2MASS\ti{}WISE & 75,754 & 0.106 & 0.105 & 0.095 & 0.099 & 0.031 & 0.024 & -3.6e-5 & 17.1\% & {3.2\%} \\ 

2MASS\ti{}SCOS &  74,866 & 0.106  &  0.106 &  0.095 &  0.097 &  0.025 &  0.022 & ~4.7e-6 &  15.0\% &  {1.6\%}\\ 

2M\ti{}WI\ti{}SC & 74,538 & 0.106  & 0.106  & 0.095  & 0.095 &  0.019 & 0.016 & -6.0e-5  &  11.2\% &  {2.4\%}\\ 

\sidehead{\bf South Galactic Cap}
2MASS\ti{}WISE & 26,692 & 0.092 & 0.092 & 0.082 & 0.085 & 0.027 & 0.021 & ~1.2e-4 & 17.1\% & {3.7\%}\\ 

2MASS\ti{}SCOS & 26,353 & 0.093 & 0.093 & 0.082 & 0.085 & 0.023 & 0.019 & ~2.1e-6 & 15.8\% & {2.3\%}\\ 

2M\ti{}WI\ti{}SC & 26,208 & 0.093 & 0.093 & 0.083 & 0.083 & 0.019 & 0.015 & -7.5e-5 & 12.1\% & {3.0\%}\\ 

\sidehead{\bf North Ecliptic Cap}

2MASS\ti{}WISE & 13,104 & 0.098 & 0.097 & 0.086 & 0.092 & 0.030 & 0.022 & -7.0e-4 & 17.4\% & {3.7\%} \\ 

2MASS\ti{}SCOS & 12,932 & 0.099 & 0.098 & 0.087 & 0.092 & 0.024 & 0.020 & -6.9e-4 & 15.9\% & {2.4\%} \\ 

2M\ti{}WI\ti{}SC & 12,844 & 0.099 & 0.099 & 0.087 & 0.089 & 0.019 & 0.015 & -5.4e-5 & 11.5\% & {2.8\%} \\ 

\sidehead{\bf All-sky spectroscopic sample}

\multicolumn{2}{l}{2MASS\ti{}WISE\ti{}SCOS}\\

~no flux limit & 300,986 & 0.094 & 0.094 & 0.083 & 0.084 & 0.019 & 0.015 & -1.5e-5 & 12.1\% & {3.0\%} \\ 

~$K_s<13.9$~mag & 207,151 & 0.078 & 0.078 & 0.070 & 0.070 & 0.017 & 0.013 & -4.9e-5 & 12.9\% & {2.9\%} \\ 

\enddata
\tablenotetext{a}{2M = 2MASS XSC (3 bands $J$,$H$,$K_s$); WI = WISE (2 bands $W1$,$W2$ {+ differential photometry}); SC = SuperCOSMOS (3 bands $B$,$R$,$I$).}
\tablenotetext{b}{In the test set.}
\tablenotetext{c}{Input (spectroscopic) redshift sample.}
\tablenotetext{d}{Output (photometric) redshift sample.}
\tablenotetext{e}{1$\sigma$ scatter between the spectroscopic and photometric redshifts, $\sigma_{\delta z}=\langle (z_{phot}-z_{spec})^2 \rangle ^{1/2}$.}
\tablenotetext{f}{Scaled median absolute deviation, $\mathrm{SMAD}(\delta z) = 1.48 \times \mathrm{med}(|\delta z-\mathrm{med}(\delta z)|)$.}
\tablenotetext{g}{Net bias of $z_{phot}$: $\langle \delta z \rangle = \langle z_{phot}-z_{spec} \rangle$.}
\tablenotetext{h}{Median of the relative  error.}
\tablenotetext{i}{{Percentage of outliers for which $|z_{phot}-z_{spec}| > 3\, \mathrm{SMAD}(\delta z)$.}\\}

\end{deluxetable*}

\begin{figure}[!b]
\centering
\includegraphics[width=0.45\textwidth]{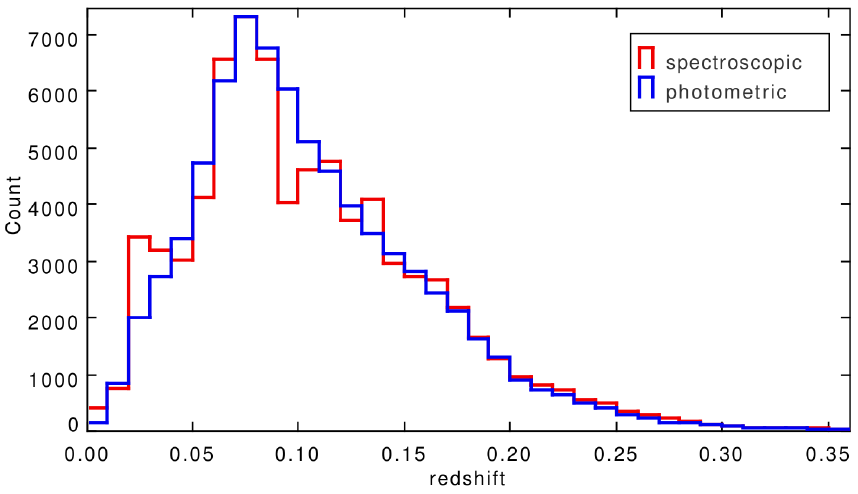}
\caption{Comparison of redshift distributions for the  test set in the North Galactic Cap: original spectroscopic   sample in red and resulting photometric redshifts in blue, where 8 bands from   2MASS\ti{}WISE\ti{}SCOS were used.}
\label{Fig: NGalCap distributions}
\end{figure}

A comparison of input spectroscopic and output photometric redshift distributions for the same test set, in the case where 8 bands were used, is shown in Figure \ref{Fig: NGalCap distributions}. The shape of $N(z)$ is  preserved to good accuracy over most of the redshift range. This is especially useful in applications where proper distributions in different redshift slices are needed, rather than the knowledge of individual galaxy redshifts, and will be taken advantage of in the forthcoming ISW analysis using this catalog together with Planck data (Steward et al., in prep.).

\subsection{South Galactic Cap}
This region has the same area as the NGCap, but it is considerably different in terms of the spectroscopic redshifts available: out of almost 113,000 '2MASS-good' sources in the region, 36\% have redshifts, the bulk of which come from the 2dFGRS, 6dFGS and SDSS (southern strips). Moreover, the median redshift of the spectroscopic sample is here considerably smaller than in the NGCap. This difference does not reflect the actual redshift distribution of 2MASS galaxies; on the contrary, the North includes Virgo and Coma clusters, whereas no such nearby overdensities exist in the SGCap.

As in the other fields, the matching rate of 2MASS-good with WISE-clean is over 99\%, it amounts to over 97\% for 2MASS\ti{}SCOS and is slightly below 97\% for 2MxWIxSC combined. Having applied all the cuts on the data, in the 8-band case we were left with over 10.8k spectroscopic redshifts for neural network training and 26.2k objects in the test set, and respectively larger numbers for the 2M\ti{}WI and 2M\ti{}SC matches. The results of applying the \ANNz\ are again summarized in Table \ref{Table 2MASS}. The statistics here are comparable to the NGCap case, despite the much smaller spectroscopic sample. Figure \ref{Fig: SGalCap 2MxWIxSC density} shows a density plot comparing spectro- and photo-$z$'s when all the 8 bands are used.

\subsection{North Ecliptic Cap}
\label{Subsec: NECap}

This field is located closer to the Galactic Plane: approximate Galactic coordinates of the of the North Ecliptic Pole are $\ell_\mathrm{NEP}=96.4^\circ$, $b_\mathrm{NEP}=29.8^\circ$. There are several reasons for choosing such a test field. First, it stretches out to the Zone of Avoidance of low Galactic latitudes, where the density of 2MASS XSC sources decreases considerably and the Galactic extinction becomes significant, even at infrared wavelengths. On the other hand, due to the WISE surveying strategy, its depth is much higher in the ecliptic caps than in the rest of the sky, reaching $5\sigma$ depths of 30 $\mu$Jy for $W1$ \citep{Jarr11}. Finally, SCOS becomes less reliable near the ZoA due to high extinction and to blends of point sources (i.e., stars) that mimic extended ones. In view of future all-sky applications, our goal was to verify the influence of these effects on the photo-$z$ estimates.

There are fewer than 98,000 '2MASS-good' sources in the region, considerably fewer than in the Galactic Caps of the same area, and their number density decreases near the Galactic plane due to lower sensitivity in the 2MASS extended source catalog, arising from stellar confusion \citep{XSCz}.  Of these, $\sim\!97\%$ sources have matches with 'WISE-clean' sample within a matching radius of $3''$, and the matching rate decreases towards lower Galactic latitudes, emphasizing the deleterious influence of stellar confusion noise. Also the percentage of matches of 2MASS XSC with SCOS sources is lower than in the two other test fields, amounting here to 95\%. The resulting 8-band 2M\ti{}WI\ti{}SC sample includes almost 91,000 sources.

The redshift coverage is relatively poor: about 20,000 galaxies have spectroscopy in this region, mainly from the SDSS and 2MRS. After all the relevant cuts, the training and test sets contained respectively $\sim$5,500 and $\sim$13,000 sources. We note that in the training and validation sets, most of the sources near the Galactic Plane were removed due to the high extinction level (for training we use only those with $EBV<0.25$, equivalent to a $B$-band extinction of $\sim$1~mag). Nevertheless, the performance of the \ANNz\ redshift estimation  -- tested on a sample disjoint with the training set, but where all the available sources with spectroscopic redshift, also those from high $EBV$ regions, were retained -- is here similar to the two previous cases and is summarized in Table \ref{Table 2MASS}. Again the additional three SCOS bands give the best improvement for the photo-$z$'s over 2MASS\ti{}WISE only. Figure \ref{Fig: NEclCap 2MxWIxSC density} shows the relevant density plot for spectroscopic and photometric redshift comparison.

\subsection{Advancing beyond 2MASS: GAMA regions}
\label{Subsec: GAMA fields}

In this Subsection we move beyond the use of 2MASS as a basic dataset for photometric redshift estimation, with the eventual aim of reaching significantly larger depths over the entire sky. We selected well-studied equatorial fields as a test-bed for future applications relying mainly on WISE data. Since both the WISE and SCOS catalogs are considerably deeper than 2MASS XSC, we matched the two former in these areas and used the currently available GAMA DR2 spectroscopic data (Liske et al.\ 2013, in prep.) to test the performance of the \ANNz\ in these fields, in view of a future all-sky WISE\ti{}SCOS photo-$z$ sample. In total, we had almost 70,000 good-quality spectroscopic redshifts released so far within the GAMA project (with $z>0.0015$ to avoid stellar contamination). As the total area of the GAMA fields that we used is 144 deg$^2$, it is clear that the GAMA number density is much higher than that of the 2MASS XSC (less than 30 sources per deg$^2$), and indeed, only about 5000 of the GAMA galaxies are also present in the 2MASS-good sample. On the other hand, the `clean' WISE data matched with the SCOS catalog within $2''$ give more than 227k pairs in the GAMA regions. Matching those additionally with the GAMA DR2 spectroscopic sources, also within $2''$, provides 63,000 `GA\ti{}WI\ti{}SC' objects. These numbers apply to those SCOS sources that have $B$ and $R$ extended-source photometry in the catalog; the $I$ band is shallower and if also used, the sample reduces to 60.2k objects. This analysis also confirms that both WISE and SCOS data are deeper than the current GAMA release and the redshift distribution of the cross-matched sample reflects the current depth of GAMA rather than of WISE\ti{}SCOS. As for the key numbers, there are over 380,000 $B,R$-selected SuperCOSMOS extended sources in the GAMA regions and over 1 million WISE objects with \ttt{w1snr}$\geq 5$ and  \ttt{w2snr}$\geq 2$.

\begin{deluxetable*}{lccccccccccc}
\tabletypesize{\footnotesize}
\tablewidth{0pt}
 \tablecolumns{13} 
\tablecaption{\label{Table GAMA}\small Statistics for the photometric redshift estimation in the GAMA regions.}

\tablehead{
 \colhead{cross-matched} & 
\colhead{number of} & 
 \multicolumn{2}{c}  {mean $\langle z \rangle$} & 
 \multicolumn{2}{c} { median $\overline{z}$} &
 \colhead{ 1$\sigma$ scatter\tablenotemark{e}} &
   \colhead{ normalized } &
 \colhead{ scaled } &
 \colhead{ net bias\tablenotemark{g} } & 
 \colhead{ median } &
 \colhead { \% of } \\
\colhead{surveys\tablenotemark{a}} &
\colhead{sources\tablenotemark{b}} &
\colhead{spec\tablenotemark{c}} & \colhead{phot\tablenotemark{d}} &
\colhead{spec\tablenotemark{c}} & \colhead{phot\tablenotemark{d}} &
\colhead{ $\sigma_{\delta z}$} &
\colhead{ $\sigma_{\delta z/(1+z)}$} &
\colhead{ MAD\tablenotemark{f} } &
\colhead{$\langle \delta z \rangle$} &
\colhead{ error\tablenotemark{h} } &
\colhead{ outliers\tablenotemark{i} }
}

\startdata

\sidehead{\bf GAMA regions}

WI\ti{}SC$_{BR}$ & 41,556 & $0.186$ &  $0.186$ & $0.175$ & $0.180$ & $0.044$ & $0.036$ &  0.034 & -2.7e-4 & $13.9\%$ &  {3.3\%} \\ 

WI\ti{}SC$_{BRI}$ & 39,536 & $0.185$ & $0.185$ & $0.174$ & $0.180$ & $0.043$ & $0.035$ & 0.034 & ~1.8e-4 & $13.7\%$ & {3.1\%} 

\enddata
\tablenotetext{a}{WI = WISE (2 bands $W1$,$W2$); SC = SuperCOSMOS (2 or 3 bands $B$,$R$,$I$).}
\tablenotetext{b}{In the test set.}
\tablenotetext{c}{Input (spectroscopic) redshift sample.}
\tablenotetext{d}{Output (photometric) redshift sample.}
\tablenotetext{e}{1$\sigma$ scatter between the spectroscopic and photometric redshifts, $\sigma_{\delta z}=\langle (z_{phot}-z_{spec})^2 \rangle ^{1/2}$.}
\tablenotetext{f}{Scaled median absolute deviation, $\mathrm{SMAD}(\delta z) = 1.48 \times \mathrm{med}(|\delta z-\mathrm{med}(\delta z)|)$.}
\tablenotetext{g}{Net bias of $z_{phot}$: $\langle \delta z \rangle = \langle z_{phot}-z_{spec} \rangle$.}
\tablenotetext{h}{Median of the relative  error.}
\tablenotetext{i}{{Percentage of outliers for which $|z_{phot}-z_{spec}| > 3\, \mathrm{SMAD}(\delta z)$.}\\}

\end{deluxetable*}

The overall sample has up to 5-band data for the photometric redshift estimation. As we show below, this is sufficient to obtain average photo-$z$ accuracy similar to the 2MASS-based analyses described in this paper, when rescaled to higher redshifts. Appropriate filtering of the data and removal of redshift outliers gave us finally a sample of almost 20,000 GA\ti{}WI\ti{}SC($BR$) sources for network training and validation (and slightly fewer with the $I$ band added). The results of applying the trained \ANNz\ to the test sets are summarized in Table \ref{Table GAMA} and illustrated in Figure \ref{Fig: GAMA WIxSC   density} comparing spectroscopic and photometric redshifts of the test sample.

\begin{figure}[b]
\centering
\includegraphics[width=0.4\textwidth]{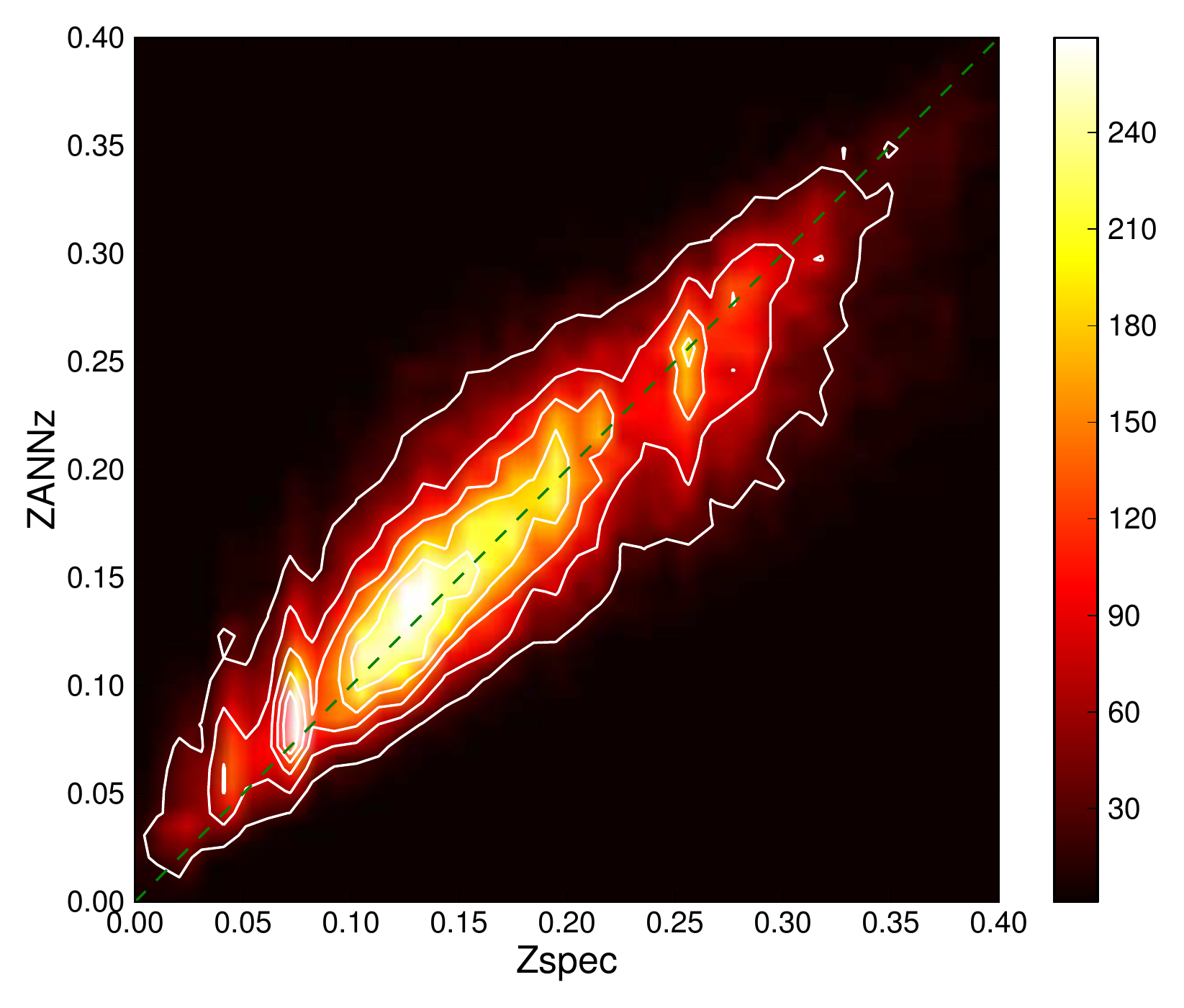}
\caption{Comparison of spectroscopic and photometric redshifts for the   GAMA regions (WISE\ti{}SCOS photometry, 4 bands; over 40,000 sources   in the test set).}
\label{Fig: GAMA WIxSC density}
\end{figure}

It is significant that even with only 4 bands used we are achieving accuracies comparable to shallower 2MASS-based samples when 5 bands (2MASS\ti{}WISE) were applied. This is very promising for future WISE\ti{}SCOS and more generally WISE\ti{}optical all-sky photo-$z$ samples, although it requires further investigation. Adding the 5th band (SCOS $I$) leads to only slight improvement and lowers the matching rate. Moreover, as the SCOS $I$-band data are reliable only to $I \lesssim19$, in our future applications using the forthcoming full GAMA~I spectroscopic dataset (which will be complete to $r\leq19.4$) and especially the GAMA~II depth of $r\leq19.8$, this additional SCOS band will have to be discarded to preserve the completeness of the WISE\ti{}SCOS sample. Similarly, the WISE $W3$ and $W4$ bands may have too low detection rates at these depths to be applicable.

\begin{figure}[b]
\centering
\includegraphics[width=0.49\textwidth ]{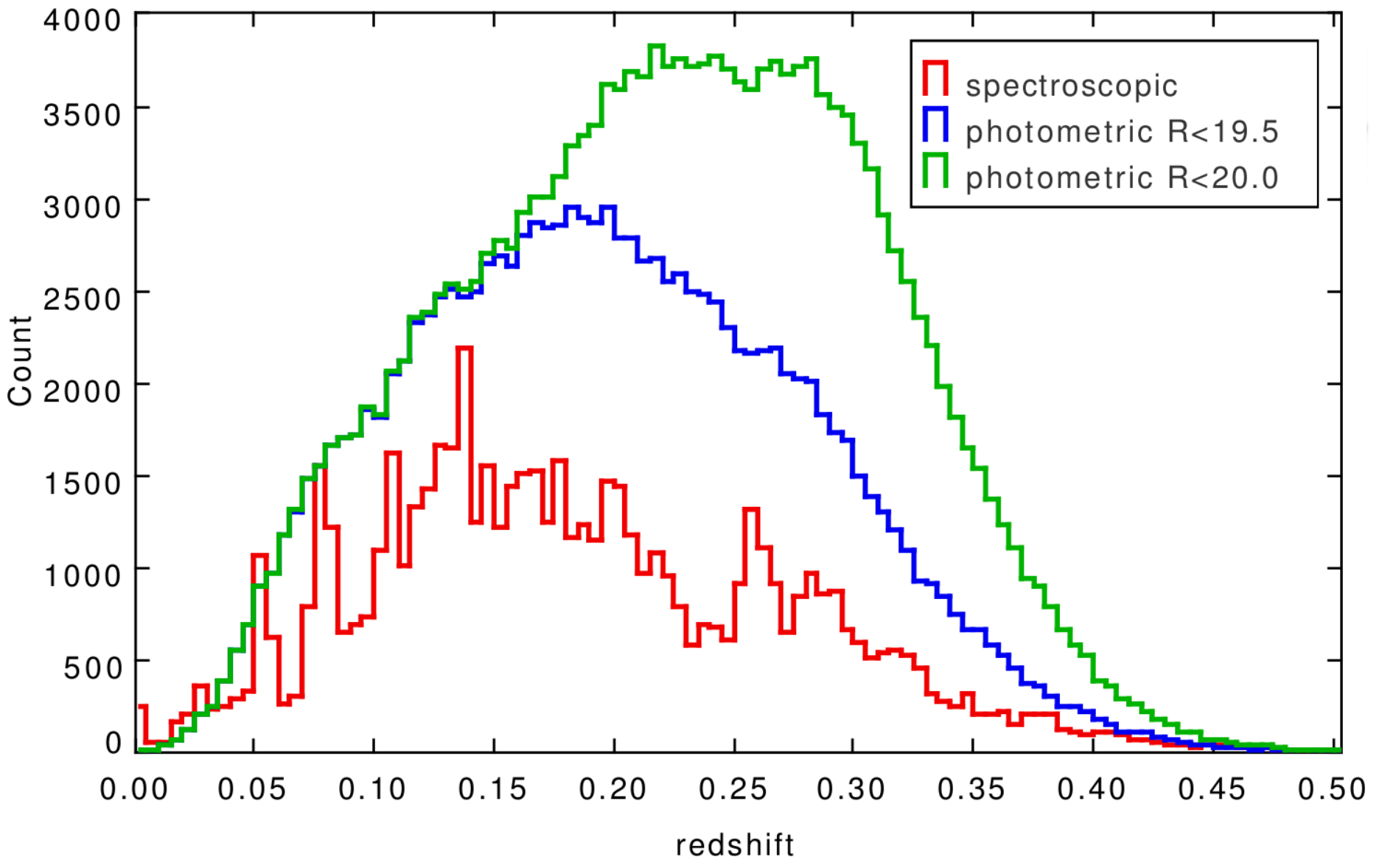} 
\caption{Redshift distribution of the spectroscopic GAMA DR2 sample (in red) and of the resulting photometric redshifts calculated for the WISE\ti{}SCOS cross-matches in GAMA areas: flux-limited at $R<19.5$ in blue and at $R<20.0$ in green.}
\label{Fig: GAMA WIxSC N of z}
\end{figure}

Having verified the performance of the \ANNz\ on the test set, we trained the neural networks on all the  GA\ti{}WI\ti{}SC cross-matches and applied them to the WISE objects in GAMA areas that paired up with an SCOS extended source. In Figure \ref{Fig: GAMA WIxSC N of z} we compare the input spectroscopic redshift distribution with the derived photometric redshifts for two flux-limited limited samples: one with $R<19.5$ and the other with $R<20.0$. These samples contain respectively 130k and 181k sources, and their median photometric redshifts are respectively 0.19 and 0.23 (compared to 0.17 in GAMA DR2). Normally it is safe to extrapolate ANNz results in this way to slightly greater depths than the calibrating spectroscopy. The full cross-matched WISE\ti{}SCOS sample reaches beyond $R=20$, but to properly calibrate photometric redshifts at these depths we would need higher redshift coverage from future GAMA releases, as the present DR2 sample is limited to $r<19.4$ in one of the fields and to $r<19.0$ in the two other. Nevertheless, a clear conclusion is that cross-matching WISE with SuperCOSMOS gives the potential to obtain an all-sky sample reaching a median redshift of at least $\sim\;$0.25.

We note finally that other redshift surveys will be of additional use for the all-sky WISE-based photometric redshift estimation, such as for instance SDSS-III. The galaxy redshift sample in the most recent Data Release of SDSS, DR10 \citep{SDSS.DR10}, has a median redshift of already $z\simeq0.3$, as it includes many higher-$z$ sources thanks to the ongoing Baryon Oscillation Spectroscopic Survey \citep[BOSS,][]{BOSS} observations. Combining current and future releases of SDSS and GAMA with legacy samples such as the \mbox{2dFGRS} should provide representative training sets for the all-sky WISE\ti{}SCOS dataset that is expected to reach up to 3 times deeper than the 2MASS-based sample presented below. One of the potential uses of this deep 3D catalog will be to select and refine sources to target for dedicated large-area multi-object spectroscopic surveys, notably the TAIPAN concept to cover the entire southern hemisphere to depths comparable to SDSS \citep{CBB13}. Among the necessary conditions for this to be achieved are the availability of a WISE extended source catalog and proper star-galaxy separation for the unresolved WISE sources (\'{S}wi\k{e}to\'{n} et al., in prep.).

\section{The Two Micron All-sky Photometric Redshift catalog}
\label{Sec: 2MPZ}

\begin{figure}[t]
\centering
\includegraphics[width=0.49\textwidth]{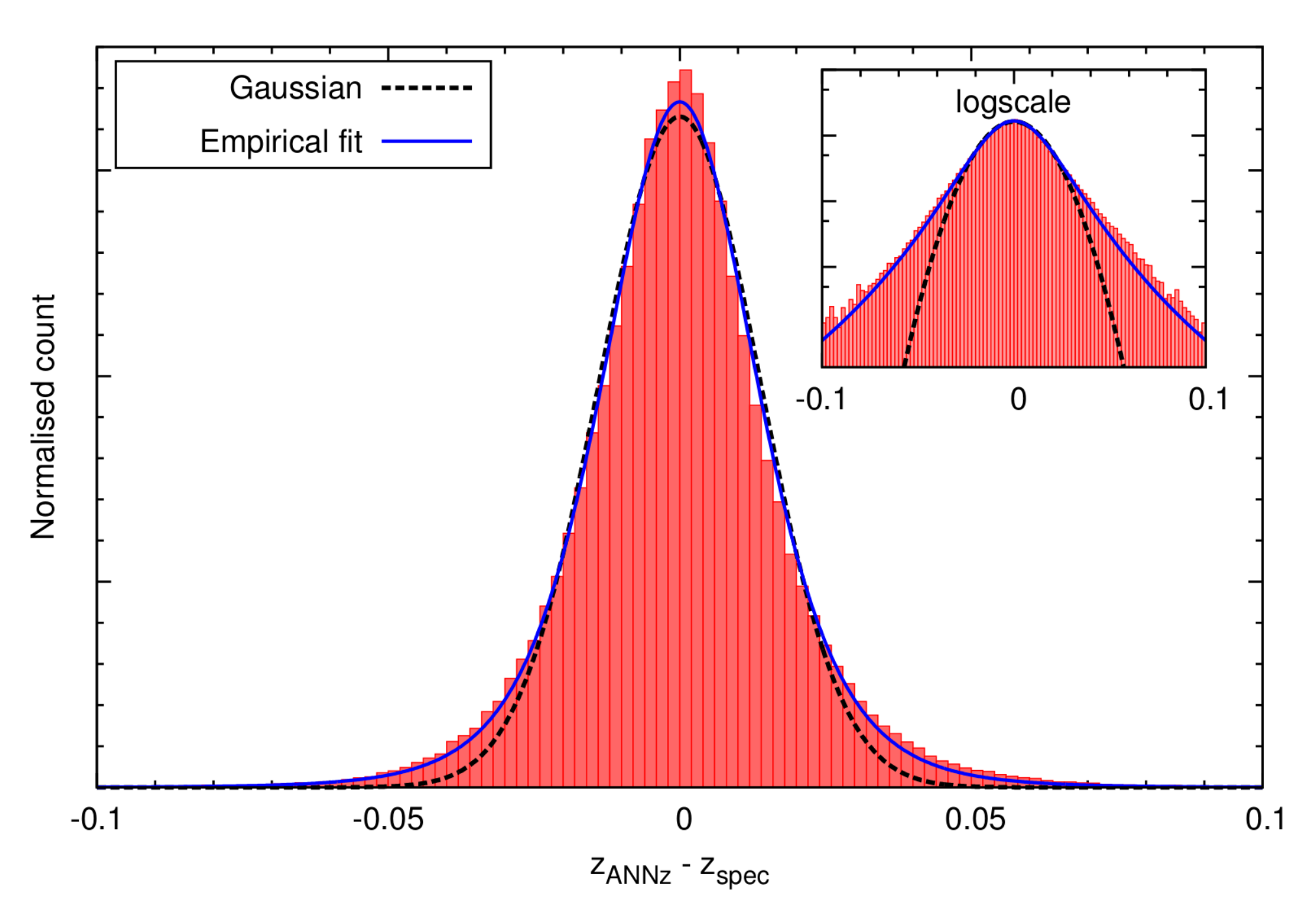}
\caption{Error distribution between the photometric and spectroscopic redshifts for the 2MASS Photometric Redshift sample (red bars). The blue solid line is the empirical fit given by Eq.\ \eqref{Eq: delta z} with $a=3.2$ and $s=0.0125$; the black dashed line shows a best-fit Gaussian with $\sigma=0.0138$.}
\label{Fig: All-sky delta z}
\end{figure}

The tests performed in the three Caps, described above, prove that we have achieved the goal of reliable photometric redshift estimation for the entire 2MASS XSC, with controlled and impressively small errors, as well as a low percentage of outliers. As we have shown, when no flux limit is imposed on the sample, the (unclipped) 1$\sigma$ scatter is below $0.02$ at a median $z\sim0.1$, scaled median absolute deviation is about $0.015$ and the median of relative error in photometric redshifts amounts to $\sim12\%$. Before generating the 2MASS Photometric Redshift catalog, we made the final checks by training and testing the neural networks on the \textit{all-sky} spectroscopic sample, described in Subsec.\ \ref{Subsec: XSCspec}. We proceeded in exactly the same way as discussed in Section \ref{Sec: Photo-z tests}, and the results are also given in Table \ref{Table 2MASS} (`All-sky spectroscopic sample'). We provide two sets of statistics: one for the full sample without any flux limit and another for sources with $K_s<13.9$, the latter being the all-sky completeness limit of 2MASS XSC. These numbers confirm our findings from the test fields, showing that flux-limiting the sample has little influence on the accuracy.

In Figure \ref{Fig: All-sky delta z} we plot the error distribution $\delta z \equiv z_\mathrm{phot} - z_\mathrm{spec}$ for the all-sky sample (red bars). Its shape is non-Gaussian in the wings and we prefer to use a more flexible fitting form instead. Namely, we chose to approximate it by
\begin{equation}\label{Eq: delta z}
N(\delta z) \propto ( 1 + \delta z^2 / 2 a s^2 )^{-a}\;,
\end{equation}
with $a\simeq 3.2$. For such a distribution, 68\% of the probability lies between $-1.2s$ and $+1.2s$ and thus defined '$1 \sigma$' is constant at $\sim0.02$ above $z=0.1$, but is smaller at lower redshifts. For the whole sample, best-fit $s=0.0125$, which is consistent with the average scatter of $\sim 0.015$. On the other hand, a Gaussian gives a much worse approximation here. For a best-fit $\sigma = 0.0138$, the central part is reasonably well matched, however the wings are underestimated. This is especially pronounced in the logscale inset of Fig.\ \ref{Fig: All-sky delta z}. Note also that if analyzed in narrow redshift bins, then at the low- and high-$z$ tail the distribution will no longer be centered on 0 nor even symmetric, which reflects the properties discussed already above and illustrated in Fig.\ \ref{Fig: NGalCap zsp.biases}.

Having checked the above properties of the photometric redshifts, we used the all-sky 2MASS\ti{}WISE\ti{}SCOS spectro-$z$ sample as training and validation sets and applied thus trained neural networks to all the 2MASS galaxies from the `good' sample, flux-limited at $K_s<13.9$, matched with both WISE and SCOS. Out of one million 2MASS galaxies present in the complete sample, almost 94\% were also identified in both the pre-filtered WISE $W1\,W2$ and SCOS $B R I$ datasets, within a matching radius of $3''$.

\begin{figure}[t]
\centering
\includegraphics[width=0.45\textwidth]{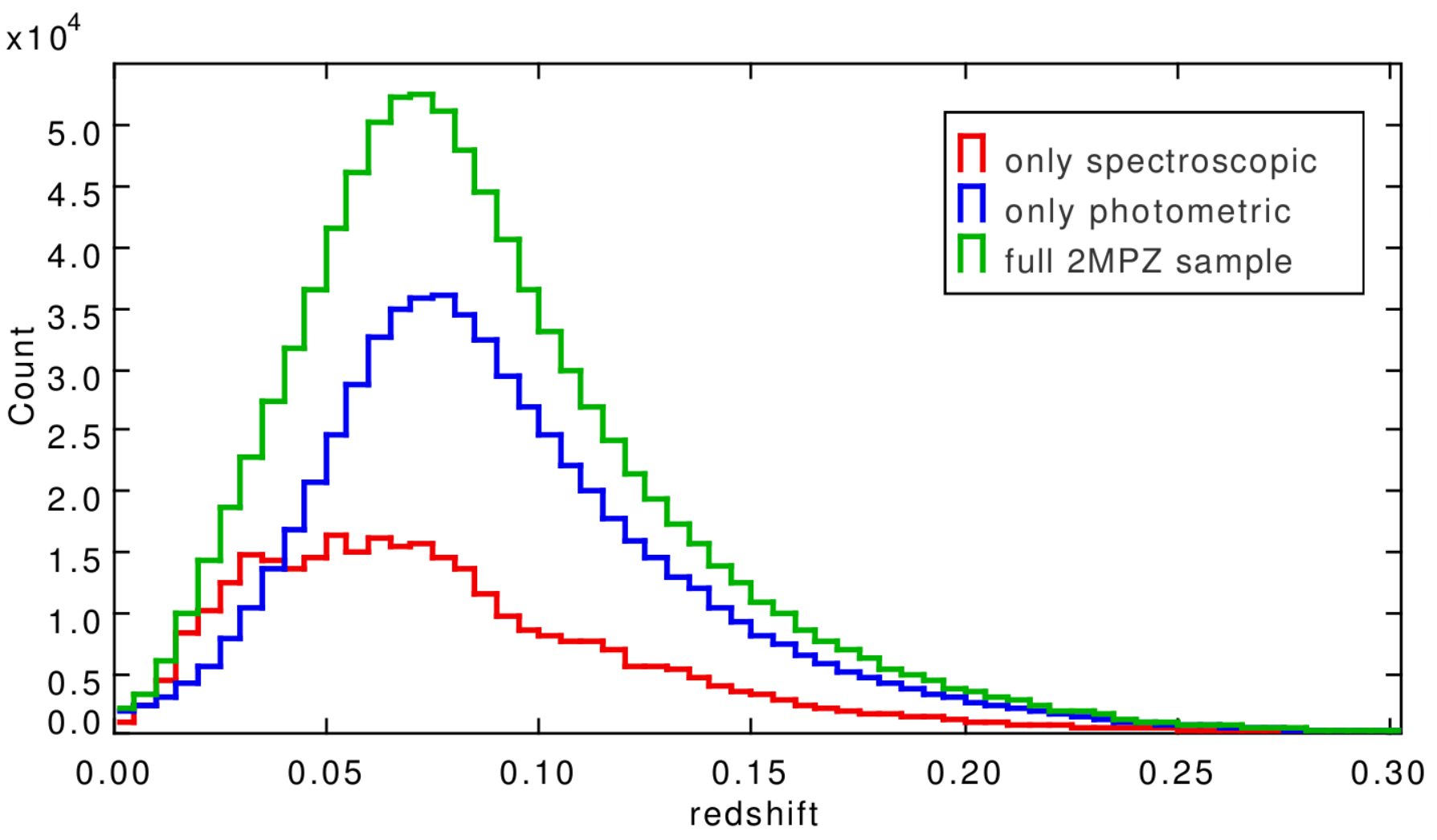}
\caption{Comparison of redshift distributions in the all-sky     2MASS Photometric Redshift catalog: sources with spectroscopic     redshifts in red, those with only photo-$z$'s in blue and the full     redshift sample in green.}
\label{Fig: 2MPZ joined-z}
\end{figure}

\begin{figure*}[!t]
\centering
\includegraphics[width=\textwidth]{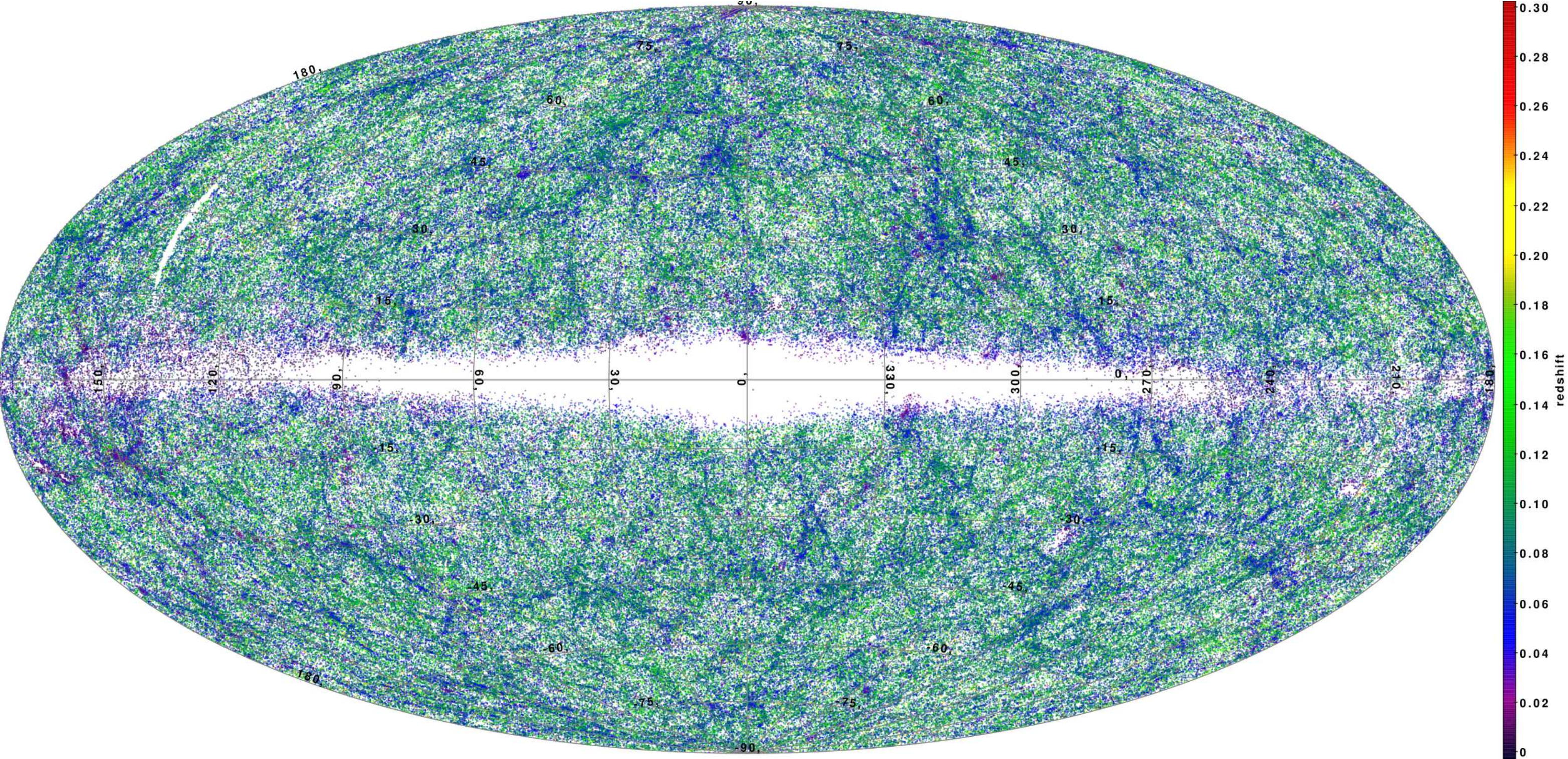}
\caption{\\ Aitoff projection in Galactic coordinates of the all-sky source distribution in the 2MASS Photometric Redshift catalog, color-coded by photometric redshift. The cosmic web is evident despite the tendency of photo-$z$'s to dilute structures.}
\label{Fig: 2MPZ Aitoff}
\end{figure*}

The small fraction of the 2MASS XSC sources that do not have   a counterpart in `clean' WISE or/and SCOS, are located mainly in the  Galactic Plane and in the regions of high extinction and stellar confusion (30\% of all the non-matches are within $|b|<10^\circ$), as well as   near the Magellanic Clouds and in the areas of WISE `torque rod   gashes' (cf.\ Figs.\ \ref{Fig: WISE all-sky} and \ref{Fig: SCOS     all-sky} and related discussion). Some of the unmatched galaxies are also   randomly scattered over the entire sky (vicinity of bright stars,   Galactic extended sources etc.), which slightly lowers the mean   number density of the cross-matched sources with respect to the   basic 2MASS XSC. The final 2MASS Photometric Redshift catalog   (2MPZ), flux limited at $K_s<13.9$, contains almost 935,000   galaxies with a mean redshift of $z=0.08$, complete over 95\% of   the sky. 

{As already discussed, in the Galactic Plane ($|b|<5^\circ$), where the   \cite{SFD} extinction maps are known to be unreliable and stellar confusion is significant, the   photometric redshifts are often biased low due to systematic errors in the   photometry. In general, the regions of   $|b|<10^\circ$ are undersampled in our 2MPZ catalog, and especially so in the spectroscopic training sets. For these reasons the Galactic plane and the above-mentioned incompleteness areas should preferably be masked out for applications requiring all-sky uniformity.} This still leaves more than 80\% of the sky comprehensively and reliably mapped in 3 dimensions with our catalog, up to distances of $\sim\;$500 Mpc.

Figure \ref{Fig: 2MPZ joined-z} compares the redshift   distributions of the 2MPZ sample: input spectroscopic redshifts in   red, derived photometric redshifts of the sources with no spectro-$z$   measurements in blue, and the distribution of the full  sample in green. Finally, Figure \ref{Fig: 2MPZ Aitoff} shows the   Aitoff projection, in Galactic coordinates, of the complete 2MPZ   catalog, color-coded by $z_\mathrm{phot}$. All the sources here have \textit{photometric} redshifts shown, even those that already had spectroscopic ones measured. The cosmic web is clearly evident with filaments, clusters and voids, despite the tendency of photo-$z$'s to dilute structures, as has been already illustrated by \cite{XSCz} where $z_\mathrm{phot}$ accuracy was considerably lower. Replacing photometric redshifts with spectroscopic ones (where available) hardly changes this overall picture.

\section{Summary and future prospects}
\label{Sec: Summary}

The main aim of this paper has been to show that by combining the largest all-sky galaxy catalogs it is possible to derive reliable and unbiased photometric redshifts over the entire sky, with reasonably small and controlled errors. We have focused mainly on the 2MASS XSC, and detailed tests on spectroscopic subsamples of this catalog have confirmed that we have met our goals. The resulting Two Micron All Sky Photometric Redshift catalog (2MPZ) contains $10^6$ galaxies, of which 2/3 have redshifts estimated only photometrically using the 8 IR and optical bands of 2MASS, WISE and SuperCOSMOS, with a 1$\sigma$ scatter of $\sigma_z = 0.015$, negligible mean bias and a very low fraction of outliers ($\sim3\%$). Having a median $z\sim0.1$, this is the largest all-sky redshift sample to date, and refines the earlier efforts of \cite{XSCz} and \cite{FP10}, as well as expands the 2M++ catalog by \cite{LH11}. Moreover, we have shown that it will be possible to obtain an all-sky 3D catalog of perhaps three times the 2MASS depth when WISE and SCOS are combined over the entire sky. For that purpose, we have used the latest redshift release of GAMA (DR2) as a test-bed, but there are many more redshift surveys available to train the photometric redshift algorithms on scales of $z \sim 0.25$. Obtaining such a WISE-based all-sky photometric redshift catalog will be our focus in the near future. This will also provide a pilot study for combining WISE with deeper large-area surveys that will be made available in the coming years, notably arising from Pan-STARRS and VISTA observations.

The applications of such three-dimensional datasets covering the whole sky will be manifold: more detailed analysis of the acceleration of the Local Group (the clustering dipole, e.g.\ \citealt{BCJM11}), of its convergence and sources of the pull \citep{KE06,Lavaux10}; bulk flow estimates on scales beyond the reach of peculiar velocities and independent of them, as well as of the kinetic Sunyaev-Zel'dovich measurements \citep{NBD11,BDN12}; measurement of the growth rate at low redshift, independent of redshift-space distortions or velocity field reconstructions \citep{NBD12}; integrated Sachs-Wolfe effect \citep{FP10} or other cross-correlations of the CMB with the LSS, such as gravitational lensing \citep{PlanckGravLens}. The future WISE-based photometric redshift catalog will enable to extend to larger volumes such tests for large-scale structure isotropy as presented by \cite{GH12}.

In principle, our catalogs will be appropriate for any applications that require uniform all-sky photometry, precise knowledge of the redshift distribution of the sample, and of the angular, radial and photometric selection function. For most of the above applications, our precision of $\sigma_z \simeq 0.015$ will be adequate. In principle, this could be further improved by using surveys with an order of magnitude more photometric bands (e.g.\ J-PAS, \citealt{J-PAS}), although there are currently no plans for such surveys to cover the entire sky.

\vspace{0.5cm}

\begin{small}

\textit{Acknowledgments.}  

We thank the Wide Field Astronomy Unit at the Institute for Astronomy, Edinburgh for archiving the 2MPZ catalog, which can be accessed at \url{http://surveys.roe.ac.uk/ssa/TWOMPZ}.

We thank the referee for useful comments improving the quality of this manuscript. We also benefited from discussions with Enzo Branchini, Adi Nusser and Ned Taylor. We are grateful to Nigel Hambly and Mike Read for providing SuperCOSMOS data in the north equatorial pole that were missing in the online database. Felipe Abdalla provided valuable help regarding the \ANNz\ code.

Special thanks to Mark Taylor for his wonderful TOPCAT software, \url{http://www.starlink.ac.uk/topcat/} \citep{TOPCAT}. Some of the results in this paper have been derived using the HEALPix package \citep{HEALPIX}, \url{http://healpix.jpl.nasa.gov}.

This publication makes use of data products from the Two Micron All Sky Survey\footnote{\url{http://www.ipac.caltech.edu/2mass/}}, Wide-field Infrared Survey Explorer\footnote{\url{http://wise.ssl.berkeley.edu/}}, SuperCOSMOS Science Archive\footnote{\url{http://surveys.roe.ac.uk/ssa/index.html}}, Sloan Digital Sky Survey\footnote{\url{http://www.sdss.org/}} and Galaxy And Mass Assembly\footnote{\url{http://www.gama-survey.org/}} catalogs.

The financial assistance of the South African National Research Foundation (NRF) towards this research is hereby acknowledged. MB was partially supported by the Polish National Science Center under contract \#UMO-2012/07/D/ST9/02785. MEC acknowledges support from the Australian Research Council (FS110200023).
 
\end{small}

\bibliography{2MPZpaper}

\end{document}